\documentclass[twocolumn]{aastex631}

\usepackage{color}
\usepackage{url}
\usepackage{hyperref}
\usepackage{xspace}
\usepackage{float}
\usepackage{soul}
\usepackage{bm}

\newcommand {\bc}{\begin {center}}
\newcommand {\ec}{\end {center}}
\newcommand {\be}{\begin {equation}}
\newcommand {\ee}{\end {equation}}
\newcommand {\beq}{\begin {eqnarray}}
\newcommand {\eeq}{\end {eqnarray}}

\newcommand {\ixpe}{{IXPE}\xspace}
\newcommand {\cen}{\mbox{Cen~X-3}\xspace}

\shorttitle{X-ray polarimetry of Cen X-3}
\shortauthors{Tsygankov et al.}
\begin{document}

%\title{IXPE observations of the first X-ray pulsar Cen X-3}

\title{The X-ray polarimetry view of the accreting pulsar Cen X-3}

\author[0000-0002-9679-0793]{Sergey~S.~Tsygankov}
\affiliation{Department of Physics and Astronomy, FI-20014 University of Turku,  Finland}
\affiliation{Space Research Institute of the Russian Academy of Sciences, Profsoyuznaya Str. 84/32, Moscow 117997, Russia}
\correspondingauthor{Sergey~S.~Tsygankov}
\email{sergey.tsygankov@utu.fi}

\author[0000-0001-8162-1105]{Victor Doroshenko}
\affiliation{Institut f\"ur Astronomie und Astrophysik, Universit\"at T\"ubingen, Sand 1, D-72076 T\"ubingen, Germany}

\author[0000-0002-0983-0049]{Juri Poutanen}
\affiliation{Department of Physics and Astronomy, FI-20014 University of Turku,  Finland}
\affiliation{Space Research Institute of the Russian Academy of Sciences, Profsoyuznaya Str. 84/32, Moscow 117997, Russia}

\author[0000-0001-9739-367X]{Jeremy Heyl}
\affiliation{Department of Physics and Astronomy, University of British Columbia, Vancouver, BC V6T 1Z1, Canada}

\author[0000-0003-2306-419X]{Alexander~A.~Mushtukov}
\affiliation{Astrophysics, Department of Physics, University of Oxford, Denys Wilkinson Building, Keble Road, Oxford OX1 3RH, UK}
\affiliation{Leiden Observatory, Leiden University, NL-2300RA Leiden, The Netherlands}

\author[0000-0002-4770-5388]{Ilaria Caiazzo}
\affiliation{TAPIR, Walter Burke Institute for Theoretical Physics, Mail Code 350-17, Caltech, Pasadena, CA 91125, USA}

\author[0000-0003-0331-3259]{Alessandro {Di Marco}}
\affiliation{INAF Istituto di Astrofisica e Planetologia Spaziali, Via del Fosso del Cavaliere 100, 00133 Roma, Italy}

\author[0000-0001-9167-2790]{Sofia V. Forsblom}
\affiliation{Department of Physics and Astronomy, FI-20014 University of Turku,  Finland}

\author[0000-0001-5848-0180]{Denis Gonz{\'a}lez-Caniulef}
\affiliation{Department of Physics and Astronomy, University of British Columbia, Vancouver, BC V6T 1Z1, Canada}

\author[0000-0001-9162-6639]{Moritz Klawin}
\affiliation{Institut f\"ur Astronomie und Astrophysik, Universit\"at T\"ubingen, Sand 1, D-72076 T\"ubingen, Germany}

\author[0000-0001-8916-4156]{Fabio {La Monaca}}
\affiliation{INAF Istituto di Astrofisica e Planetologia Spaziali, Via del Fosso del Cavaliere 100, 00133 Roma, Italy}

\author[0000-0002-0380-0041]{Christian Malacaria}
\affiliation{International Space Science Institute, Hallerstrasse 
6, 3012 Bern, Switzerland}

\author[0000-0002-6492-1293]{Herman L. Marshall}
\affiliation{MIT Kavli Institute for Astrophysics and Space Research, Massachusetts Institute of Technology, 77 Massachusetts Avenue, Cambridge, MA 02139, USA}

%internal referee
\author[0000-0003-3331-3794]{Fabio Muleri}
\affiliation{INAF Istituto di Astrofisica e Planetologia Spaziali, Via del Fosso del Cavaliere 100, 00133 Roma, Italy}

\author[0000-0002-0940-6563]{Mason Ng}
\affiliation{MIT Kavli Institute for Astrophysics and Space Research, Massachusetts Institute of Technology, 77 Massachusetts Avenue, Cambridge, MA 02139, USA}

\author[0000-0003-3733-7267]{Valery F. Suleimanov}
\affiliation{Institut f\"ur Astronomie und Astrophysik, Universit\"at T\"ubingen, Sand 1, D-72076 T\"ubingen, Germany}

\author[0000-0002-2764-7192]{Rashid A. Sunyaev}
\affiliation{Max Planck Institute for Astrophysics, Karl-Schwarzschild-Str 1, D-85741 Garching, Germany}
\affiliation{Space Research Institute of the Russian Academy of Sciences, Profsoyuznaya Str. 84/32, Moscow 117997, Russia}

%internal referee
\author[0000-0003-3977-8760]{Roberto Turolla}
\affiliation{Dipartimento di Fisica e Astronomia, Universit\'{a} degli Studi di Padova, Via Marzolo 8, 35131 Padova, Italy}
\affiliation{Mullard Space Science Laboratory, University College London, Holmbury St Mary, Dorking, Surrey RH5 6NT, UK}

\author[0000-0002-3777-6182]{Iv\'an Agudo}
\affiliation{Instituto de Astrof\'{i}sicade Andaluc\'{i}a -- CSIC, Glorieta de la Astronom\'{i}a s/n, 18008 Granada, Spain}
\author[0000-0002-5037-9034]{Lucio A. Antonelli}
\affiliation{INAF Osservatorio Astronomico di Roma, Via Frascati 33, 00040 Monte Porzio Catone (RM), Italy}
\affiliation{Space Science Data Center, Agenzia Spaziale Italiana, Via del Politecnico snc, 00133 Roma, Italy}
\author[0000-0002-4576-9337]{Matteo Bachetti}
\affiliation{INAF Osservatorio Astronomico di Cagliari, Via della Scienza 5, 09047 Selargius (CA), Italy}
\author[0000-0002-9785-7726]{Luca Baldini}
\affiliation{Istituto Nazionale di Fisica Nucleare, Sezione di Pisa, Largo B. Pontecorvo 3, 56127 Pisa, Italy}
\affiliation{Dipartimento di Fisica, Universit\`{a} di Pisa, Largo B. Pontecorvo 3, 56127 Pisa, Italy}
\author{Wayne H. Baumgartner}
\affiliation{NASA Marshall Space Flight Center, Huntsville, AL 35812, USA}
\author[0000-0002-2469-7063]{Ronaldo Bellazzini}
\affiliation{Istituto Nazionale di Fisica Nucleare, Sezione di Pisa, Largo B. Pontecorvo 3, 56127 Pisa, Italy}
\author[0000-0002-4622-4240]{Stefano Bianchi}
\affiliation{Dipartimento di Matematica e Fisica, Universit\`{a} degli Studi Roma Tre, Via della Vasca Navale 84, 00146 Roma, Italy}
\author{Stephen D. Bongiorno}
\affiliation{NASA Marshall Space Flight Center, Huntsville, AL 35812, USA}
\author[0000-0002-4264-1215]{Raffaella Bonino}
\affiliation{Istituto Nazionale di Fisica Nucleare, Sezione di Torino, Via Pietro Giuria 1, 10125 Torino, Italy}
\affiliation{Dipartimento di Fisica, Universit\`{a} degli Studi di Torino, Via Pietro Giuria 1, 10125 Torino, Italy}
\author[0000-0002-9460-1821]{Alessandro Brez}
\affiliation{Istituto Nazionale di Fisica Nucleare, Sezione di Pisa, Largo B. Pontecorvo 3, 56127 Pisa, Italy}
\author[0000-0002-8848-1392]{Niccol\`{o} Bucciantini}
\affiliation{INAF Osservatorio Astrofisico di Arcetri, Largo Enrico Fermi 5, 50125 Firenze, Italy}
\affiliation{Dipartimento di Fisica e Astronomia, Universit\`{a} degli Studi di Firenze, Via Sansone 1, 50019 Sesto Fiorentino (FI), Italy}
\affiliation{Istituto Nazionale di Fisica Nucleare, Sezione di Firenze, Via Sansone 1, 50019 Sesto Fiorentino (FI), Italy}
\author[0000-0002-6384-3027]{Fiamma Capitanio}
\affiliation{INAF Istituto di Astrofisica e Planetologia Spaziali, Via del Fosso del Cavaliere 100, 00133 Roma, Italy}
\author[0000-0003-1111-4292]{Simone Castellano}
\affiliation{Istituto Nazionale di Fisica Nucleare, Sezione di Pisa, Largo B. Pontecorvo 3, 56127 Pisa, Italy}
\author[0000-0001-7150-9638]{Elisabetta Cavazzuti}
\affiliation{Agenzia Spaziale Italiana, Via del Politecnico snc, 00133 Roma, Italy}
\author[0000-0002-0712-2479]{Stefano Ciprini}
\affiliation{Istituto Nazionale di Fisica Nucleare, Sezione di Roma ``Tor Vergata'', Via della Ricerca Scientifica 1, 00133 Roma, Italy}
\affiliation{Space Science Data Center, Agenzia Spaziale Italiana, Via del Politecnico snc, 00133 Roma, Italy}
\author[0000-0003-4925-8523]{Enrico Costa}
\affiliation{INAF Istituto di Astrofisica e Planetologia Spaziali, Via del Fosso del Cavaliere 100, 00133 Roma, Italy}
\author[0000-0001-5668-6863]{Alessandra De Rosa}
\affiliation{INAF Istituto di Astrofisica e Planetologia Spaziali, Via del Fosso del Cavaliere 100, 00133 Roma, Italy}
\author[0000-0002-3013-6334]{Ettore Del Monte}
\affiliation{INAF Istituto di Astrofisica e Planetologia Spaziali, Via del Fosso del Cavaliere 100, 00133 Roma, Italy}
\author[0000-0002-5614-5028]{Laura Di Gesu}
\affiliation{Agenzia Spaziale Italiana, Via del Politecnico snc, 00133 Roma, Italy}
\author[0000-0002-7574-1298]{Niccol\`{o} Di Lalla}
\affiliation{Department of Physics and Kavli Institute for Particle Astrophysics and Cosmology, Stanford University, Stanford, California 94305, USA}
\author[0000-0002-4700-4549]{Immacolata Donnarumma}
\affiliation{Agenzia Spaziale Italiana, Via del Politecnico snc, 00133 Roma, Italy}
\author[0000-0003-0079-1239]{Michal Dov\v{c}iak}
\affiliation{Astronomical Institute of the Czech Academy of Sciences, Bo\v{c}n\'{i} II 1401/1, 14100 Praha 4, Czech Republic}
\author[0000-0003-4420-2838]{Steven R. Ehlert}
\affiliation{NASA Marshall Space Flight Center, Huntsville, AL 35812, USA}
\author[0000-0003-1244-3100]{Teruaki Enoto}
\affiliation{RIKEN Cluster for Pioneering Research, 2-1 Hirosawa, Wako, Saitama 351-0198, Japan}
\author[0000-0001-6096-6710]{Yuri Evangelista}
\affiliation{INAF Istituto di Astrofisica e Planetologia Spaziali, Via del Fosso del Cavaliere 100, 00133 Roma, Italy}
\author[0000-0003-1533-0283]{Sergio Fabiani}
\affiliation{INAF Istituto di Astrofisica e Planetologia Spaziali, Via del Fosso del Cavaliere 100, 00133 Roma, Italy}
\author[0000-0003-1074-8605]{Riccardo Ferrazzoli}
\affiliation{INAF Istituto di Astrofisica e Planetologia Spaziali, Via del Fosso del Cavaliere 100, 00133 Roma, Italy}
\author[0000-0003-3828-2448]{Javier A. Garcia}
\affiliation{California Institute of Technology, Pasadena, CA 91125, USA}
\author[0000-0002-5881-2445]{Shuichi Gunji}
\affiliation{Yamagata University,1-4-12 Kojirakawa-machi, Yamagata-shi 990-8560, Japan}
\author{Kiyoshi Hayashida}
\altaffiliation{Deceased}
\affiliation{Osaka University, 1-1 Yamadaoka, Suita, Osaka 565-0871, Japan}
\author[0000-0002-0207-9010]{Wataru Iwakiri}
\affiliation{Department of Physics, Faculty of Science and Engineering, Chuo University, 1-13-27 Kasuga, Bunkyo-ku, Tokyo 112-8551, Japan}
\author[0000-0001-9522-5453]{Svetlana G. Jorstad}
\affiliation{Institute for Astrophysical Research, Boston University, 725 Commonwealth Avenue, Boston, MA 02215, USA}
\affiliation{Department of Astrophysics, St. Petersburg State University, Universitetsky pr. 28, Petrodvoretz, 198504 St. Petersburg, Russia}
\author[0000-0002-5760-0459]{Vladimir Karas}
\affiliation{Astronomical Institute of the Czech Academy of Sciences, Bo\v{c}n\'{i} II 1401/1, 14100 Praha 4, Czech Republic}
\author{Takao Kitaguchi}
\affiliation{RIKEN Cluster for Pioneering Research, 2-1 Hirosawa, Wako, Saitama 351-0198, Japan}
\author[0000-0002-0110-6136]{Jeffery J. Kolodziejczak}
\affiliation{NASA Marshall Space Flight Center, Huntsville, AL 35812, USA}
\author[0000-0002-1084-6507]{Henric Krawczynski}
\affiliation{Physics Department and McDonnell Center for the Space Sciences, Washington University in St. Louis, St. Louis, MO 63130, USA}
\author[0000-0002-0984-1856]{Luca Latronico}
\affiliation{Istituto Nazionale di Fisica Nucleare, Sezione di Torino, Via Pietro Giuria 1, 10125 Torino, Italy}
\author[0000-0001-9200-4006]{Ioannis Liodakis}
\affiliation{Finnish Centre for Astronomy with ESO,  20014 University of Turku, Finland}
\author[0000-0002-0698-4421]{Simone Maldera}
\affiliation{Istituto Nazionale di Fisica Nucleare, Sezione di Torino, Via Pietro Giuria 1, 10125 Torino, Italy}
\author[0000-0002-0998-4953]{Alberto Manfreda}  
\affiliation{Istituto Nazionale di Fisica Nucleare, Sezione di Pisa, Largo B. Pontecorvo 3, 56127 Pisa, Italy}
\author[0000-0003-4952-0835]{Fr\'{e}d\'{e}ric Marin}
\affiliation{Universit\'{e} de Strasbourg, CNRS, Observatoire Astronomique de Strasbourg, UMR 7550, 67000 Strasbourg, France}
\author[0000-0002-2055-4946]{Andrea Marinucci}
\affiliation{Agenzia Spaziale Italiana, Via del Politecnico snc, 00133 Roma, Italy}
\author[0000-0001-7396-3332]{Alan P. Marscher}
\affiliation{Institute for Astrophysical Research, Boston University, 725 Commonwealth Avenue, Boston, MA 02215, USA}
\author[0000-0002-2152-0916]{Giorgio Matt}
\affiliation{Dipartimento di Matematica e Fisica, Universit\`{a} degli Studi Roma Tre, Via della Vasca Navale 84, 00146 Roma, Italy}
\author{Ikuyuki Mitsuishi}
\affiliation{Graduate School of Science, Division of Particle and Astrophysical Science, Nagoya University, Furo-cho, Chikusa-ku, Nagoya, Aichi 464-8602, Japan}
\author[0000-0001-7263-0296]{Tsunefumi Mizuno}
\affiliation{Hiroshima Astrophysical Science Center, Hiroshima University, 1-3-1 Kagamiyama, Higashi-Hiroshima, Hiroshima 739-8526, Japan}
\author[0000-0002-5847-2612]{C.-Y. Ng}
\affiliation{Department of Physics, University of Hong Kong, Pokfulam, Hong Kong}
\author[0000-0002-1868-8056]{Stephen L. O'Dell}
\affiliation{NASA Marshall Space Flight Center, Huntsville, AL 35812, USA}
\author[0000-0002-5448-7577]{Nicola Omodei}
\affiliation{Department of Physics and Kavli Institute for Particle Astrophysics and Cosmology, Stanford University, Stanford, California 94305, USA}
\author[0000-0001-6194-4601]{Chiara Oppedisano}
\affiliation{Istituto Nazionale di Fisica Nucleare, Sezione di Torino, Via Pietro Giuria 1, 10125 Torino, Italy}
\author[0000-0001-6289-7413]{Alessandro Papitto}
\affiliation{INAF Osservatorio Astronomico di Roma, Via Frascati 33, 00040 Monte Porzio Catone (RM), Italy}
\author[0000-0002-7481-5259]{George G. Pavlov}
\affiliation{Department of Astronomy and Astrophysics, Pennsylvania State University, University Park, PA 16801, USA}
\author[0000-0001-6292-1911]{Abel L. Peirson}
\affiliation{Department of Physics and Kavli Institute for Particle Astrophysics and Cosmology, Stanford University, Stanford, California 94305, USA}
\author[0000-0003-3613-4409]{Matteo Perri}
\affiliation{Space Science Data Center, Agenzia Spaziale Italiana, Via del Politecnico snc, 00133 Roma, Italy}
\affiliation{INAF Osservatorio Astronomico di Roma, Via Frascati 33, 00040 Monte Porzio Catone (RM), Italy}
\author[0000-0003-1790-8018]{Melissa Pesce-Rollins}
\affiliation{Istituto Nazionale di Fisica Nucleare, Sezione di Pisa, Largo B. Pontecorvo 3, 56127 Pisa, Italy}
\author[0000-0001-6061-3480]{Pierre-Olivier Petrucci}
\affiliation{Universit\'{e} Grenoble Alpes, CNRS, IPAG, 38000 Grenoble, France}
\author[0000-0001-7397-8091]{Maura Pilia}
\affiliation{INAF Osservatorio Astronomico di Cagliari, Via della Scienza 5, 09047 Selargius (CA), Italy}
\author[0000-0001-5902-3731]{Andrea Possenti}
\affiliation{INAF Osservatorio Astronomico di Cagliari, Via della Scienza 5, 09047 Selargius (CA), Italy}
\author[0000-0002-2734-7835]{Simonetta Puccetti}
\affiliation{Space Science Data Center, Agenzia Spaziale Italiana, Via del Politecnico snc, 00133 Roma, Italy}
\author{Brian D. Ramsey}
\affiliation{NASA Marshall Space Flight Center, Huntsville, AL 35812, USA}
\author[0000-0002-9774-0560]{John Rankin}
\affiliation{INAF Istituto di Astrofisica e Planetologia Spaziali, Via del Fosso del Cavaliere 100, 00133 Roma, Italy}
\author[0000-0003-0411-4243]{Ajay Ratheesh}
\affiliation{INAF Istituto di Astrofisica e Planetologia Spaziali, Via del Fosso del Cavaliere 100, 00133 Roma, Italy}
\author[0000-0001-6711-3286]{Roger W. Romani}
\affiliation{Department of Physics and Kavli Institute for Particle Astrophysics and Cosmology, Stanford University, Stanford, California 94305, USA}
\author[0000-0001-5676-6214]{Carmelo Sgr\`{o}}
\affiliation{Istituto Nazionale di Fisica Nucleare, Sezione di Pisa, Largo B. Pontecorvo 3, 56127 Pisa, Italy}
\author[0000-0002-6986-6756]{Patrick Slane}
\affiliation{Center for Astrophysics, Harvard \& Smithsonian, 60 Garden St, Cambridge, MA 02138, USA}
\author[0000-0001-8916-4156]{Paolo Soffitta}
\affiliation{INAF Istituto di Astrofisica e Planetologia Spaziali, Via del Fosso del Cavaliere 100, 00133 Roma, Italy}
\author[0000-0003-0802-3453]{Gloria Spandre}
\affiliation{Istituto Nazionale di Fisica Nucleare, Sezione di Pisa, Largo B. Pontecorvo 3, 56127 Pisa, Italy}
\author[0000-0002-8801-6263]{Toru Tamagawa}
\affiliation{RIKEN Cluster for Pioneering Research, 2-1 Hirosawa, Wako, Saitama 351-0198, Japan}
\author[0000-0003-0256-0995]{Fabrizio Tavecchio}
\affiliation{INAF Osservatorio Astronomico di Brera, via E. Bianchi 46, 23807 Merate (LC), Italy}
\author[0000-0002-1768-618X]{Roberto Taverna}
\affiliation{Dipartimento di Fisica e Astronomia, Universit\`{a} degli Studi di Padova, Via Marzolo 8, 35131 Padova, Italy}
\author{Yuzuru Tawara}
\affiliation{Graduate School of Science, Division of Particle and Astrophysical Science, Nagoya University, Furo-cho, Chikusa-ku, Nagoya, Aichi 464-8602, Japan}
\author{Allyn F. Tennant}
\affiliation{NASA Marshall Space Flight Center, Huntsville, AL 35812, USA}
\author{Nicolas E. Thomas}
\affiliation{NASA Marshall Space Flight Center, Huntsville, AL 35812, USA}
\author[0000-0002-6562-8654]{Francesco Tombesi}
\affiliation{Dipartimento di Fisica, Universit\`{a} degli Studi di Roma ``Tor Vergata'', Via della Ricerca Scientifica 1, 00133 Roma, Italy}
\affiliation{Istituto Nazionale di Fisica Nucleare, Sezione di Roma ``Tor Vergata'', Via della Ricerca Scientifica 1, 00133 Roma, Italy}
\affiliation{Department of Astronomy, University of Maryland, College Park, Maryland 20742, USA}
\author[0000-0002-3180-6002]{Alessio Trois}
\affiliation{INAF Osservatorio Astronomico di Cagliari, Via della Scienza 5, 09047 Selargius (CA), Italy}
\author[0000-0002-4708-4219]{Jacco Vink}
\affiliation{Anton Pannekoek Institute for Astronomy \& GRAPPA, University of Amsterdam, Science Park 904, 1098 XH Amsterdam, The Netherlands}
\author[0000-0002-5270-4240]{Martin C. Weisskopf}
\affiliation{NASA Marshall Space Flight Center, Huntsville, AL 35812, USA}
\author[0000-0002-7568-8765]{Kinwah Wu}
\affiliation{Mullard Space Science Laboratory, University College London, Holmbury St Mary, Dorking, Surrey RH5 6NT, UK}
\author[0000-0002-0105-5826]{Fei Xie}
\affiliation{Guangxi Key Laboratory for Relativistic Astrophysics, School of Physical Science and Technology, Guangxi University, Nanning 530004, China}
\affiliation{INAF Istituto di Astrofisica e Planetologia Spaziali, Via del Fosso del Cavaliere 100, 00133 Roma, Italy}
\author[0000-0001-5326-880X]{Silvia Zane}
\affiliation{Mullard Space Science Laboratory, University College London, Holmbury St Mary, Dorking, Surrey RH5 6NT, UK}

\collaboration{98}{(IXPE Collaboration)}



\begin{abstract}
\cen is the first X-ray pulsar discovered 50 years ago.
Radiation from such objects is expected to be highly polarized due to birefringence of plasma and vacuum associated with propagation of photons in presence of the strong magnetic field. 
Here we present results of the  observations of \cen performed with the  {Imaging X-ray Polarimetry Explorer}. The source exhibited significant flux variability and was observed in two states different by a factor of $\sim20$ in flux. In the low-luminosity state no significant polarization was found either in pulse phase-averaged (with the 3$\sigma$ upper limit of 12\%) or phase-resolved data (the  3$\sigma$ upper limits are 20--30\%).
In the bright state the polarization degree of 5.8$\pm$0.3\% and polarization angle of $49\fdg6\pm1\fdg5$ with significance of about 20$\sigma$ was measured from the spectro-polarimetric analysis of the phase-averaged data. 
The phase-resolved analysis showed a significant anti-correlation between the flux and the polarization degree as well as strong variations of the polarization angle. 
The fit with the rotating vector model indicates a position angle of the pulsar spin axis of about 49\degr\ and a magnetic obliquity of 17\degr. 
The detected relatively low polarization can be explained if the upper layers of the neutron star surface are overheated by the accreted matter and the conversion of the polarization modes occurs within the transition region between the upper hot layer and a cooler underlying atmosphere. 
A fraction of polarization signal can also be produced by reflection of radiation from the neutron star surface and the accretion curtain.
\end{abstract}



\section{Introduction} 
\label{sec:intro}

Understanding the interaction of astrophysical plasmas with ultra-strong magnetic and radiation fields in the vicinity of neutron stars (NSs) is a stumbling block of modern astrophysics which cannot be addressed in terrestrial labs. 
One of the immediate consequences of this interaction is funneling of the accreting matter towards two small polar regions on the NS surface. 
Release of the kinetic energy of matter results in strong emission peaked in the X-ray energy band and the consequent appearance of an X-ray pulsar \citep[XRP; see][for a recent review]{2022arXiv220414185M}. 

The extreme magnetic field at the surface of a NS influences many aspects of plasma physics on the level of elementary processes determining interaction of radiation and matter \citep{2006RPPh...69.2631H}. The emerging radiation, therefore, allows to deduce the configuration of the emission regions and even to detect specific effects of quantum electrodynamics, shedding light on physics of the interaction of radiation and matter under conditions of extremely strong magnetic fields.

According to current theoretical models, emission of XRPs is expected to be strongly polarized (up to 80\%, see, e.g., \citealt{1988ApJ...324.1056M,2021MNRAS.501..109C}).
Many authors addressed the polarization properties of XRPs \citep{1981ApJ...251..278N,1981ApJ...251..288N,1982Ap&SS..86..249K,1985ApJ...298..147M,1985ApJ...299..138M,1986PASJ...38..751K,1987PASJ...39..781K} with the most recent models covering low \citep{2021MNRAS.503.5193M,2021A&A...651A..12S} and high \citep{2021MNRAS.501..109C} mass accretion rates with thin slab and accretion column geometries, respectively.
A relatively low polarization degree (PD) remains possible at low mass accretion rates due to the inverse temperature profile in the atmosphere of an accreting NS \citep{2019MNRAS.483..599G,2021MNRAS.503.5193M}. Verification of different model predictions requires sensitive polarimetric observations.

Until recently, only a couple of attempts have been made to detect the polarimetric signal from XRPs with different instruments.  
A search for X-ray polarization in the XRPs Cen X-3 and Her X-1 was performed in the {OSO-8} polarimeter data collected in 1975 \citep{1979ApJ...232..248S}. However, no significant polarization was detected either in pulse phase-averaged or phase-resolved data for both sources. 
Recently,  the balloon-borne  hard X-ray polarimeter X-Calibur observed GX~301$-$2 in the 15--35 keV energy range, but, again, no significant polarization signal was detected  \citep{2020ApJ...891...70A}. 

The first highly sensitive space X-ray polarimeter, the {Imaging X-ray Polarimeter Explorer} \citep[\ixpe,][]{Weisskopf2022}, was launched on 2021 December 9.
The high-quality data allowed \cite{Doroshenko2022} to discover a significant polarization signal from \hbox{Her~X-1} both in the phase-averaged and phase-resolved data with PD varying from $\sim5$\% to $\sim15$\% over the rotational phase.

\cen was also observed by \ixpe as one of the first science targets. Coherent pulsations from the source with period of about 4.87~s were discovered by the first X-ray space observatory {Uhuru} \citep{1971ApJ...167L..67G}. The binary nature of \cen was established by \cite{1972ApJ...172L..79S}, who detected the occultations of the X-ray source by a massive optical companion, as well as Doppler shifts of the apparent spin frequency, with an orbital period of $\sim2.09$~d. Today it is known that the system consists of a NS of mass $M_{\rm NS}=1.2\pm0.2\,M_{\odot}$ in an almost circular orbit ($e < 0.0016$, orbital separation  $a\approx 19R_{\odot}$) around the O6–8 II-III supergiant V779~Cen of mass $M_{\rm O}=20.5\pm0.7\, M_{\odot}$ and radius $R_{\rm O} \sim12R_{\odot}$  \citep{1974ApJ...192L.135K,1999MNRAS.307..357A,2010MNRAS.401.1532R}.  Assuming that the optical component fills its Roche lobe, and that the system is in synchronous rotation, \cite{1999MNRAS.307..357A} determined the orbital inclination of the system to be $i_{\rm orb}=70\fdg2\pm2\fdg7$. 

The primary channel of mass transfer in the system is Roche-lobe overflow, resulting in the formation of an accretion disk around the NS \citep{1986A&A...154...77T}, although stellar wind from the companion is also clearly present and may contribute to accretion. The distance to \cen, $d=6.4^{+1.0}_{-1.4}$~kpc, was recently refined based on \textit{Gaia} data \citep{2021MNRAS.502.5455A}.

The magnetic field strength of the NS is known from the presence of a cyclotron resonant scattering feature in the source spectrum at an energy around 30 keV first discovered with {Ginga} \citep{1992ApJ...396..147N}. It was later confirmed by {BeppoSAX}, {RXTE}, {Suzaku}, and {NuSTAR} \citep{1998A&A...340L..55S,1999hxra.conf...25H,2021MNRAS.500.3454T}. Moreover, it was shown that while the line energy does not depend on the source luminosity \citep{2021MNRAS.500.3454T} and on time \citep{2019MNRAS.484.3797J}, its parameters substantially change with the rotational phase  \citep[see, e.g.,][]{2000ApJ...530..429B,2008ApJ...675.1487S} as expected given the strong angular dependence of the scattering cross-section in the magnetic field.

%%%%%%%%%%%%%%%%%%%%%%%%%%
\begin{figure*}
\centering
\includegraphics[height=0.43\linewidth]{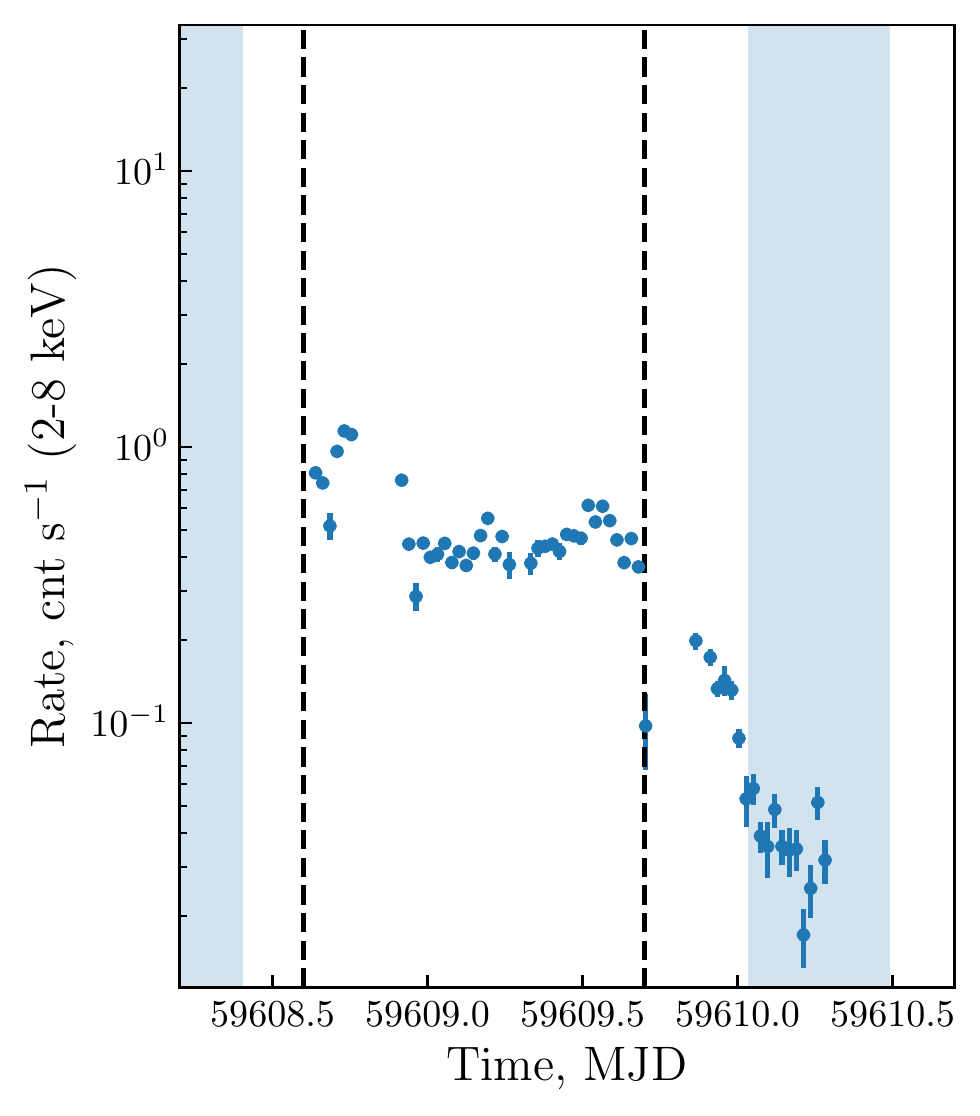}
\includegraphics[height=0.43\linewidth]{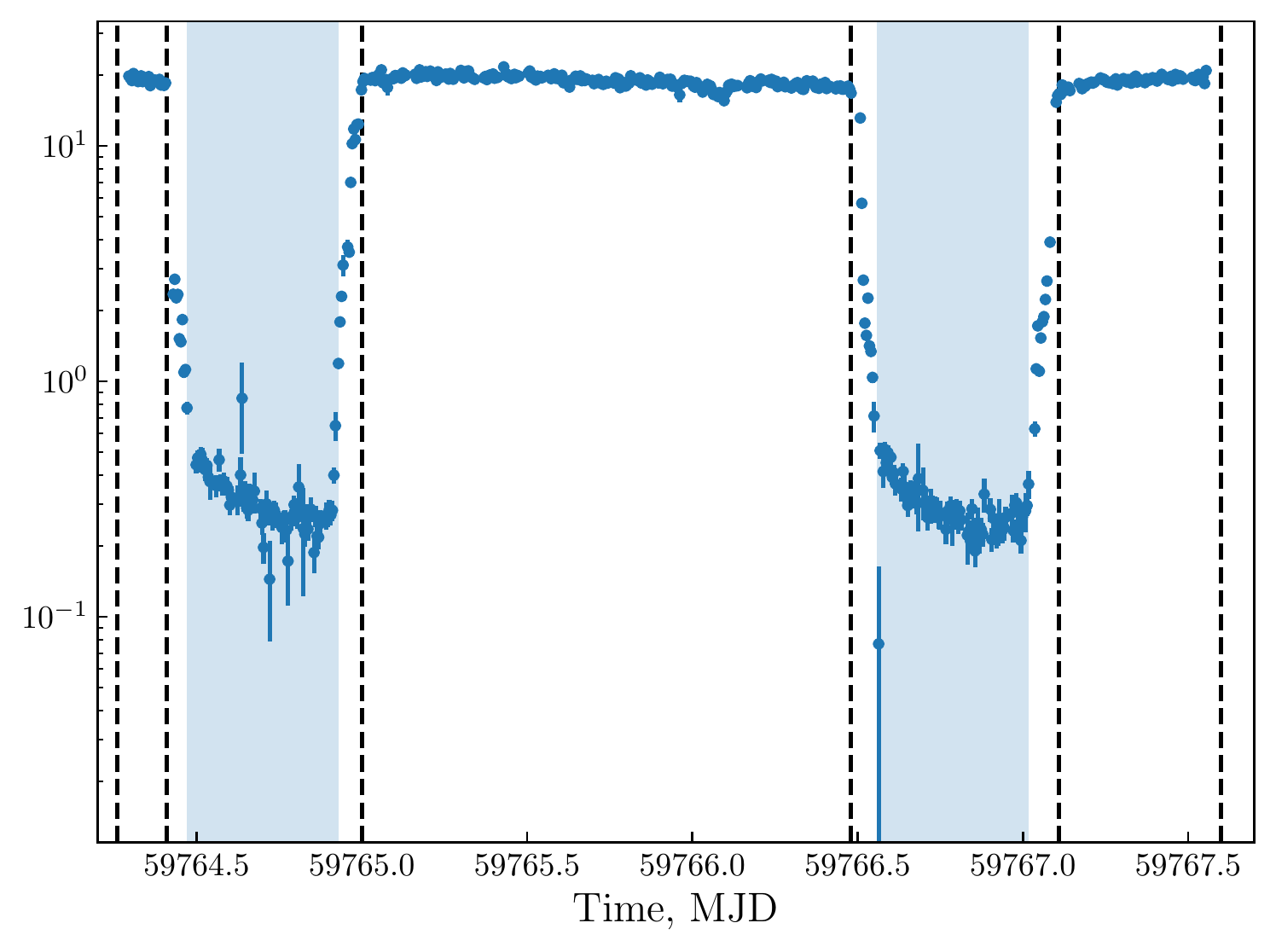}
\caption{Background corrected light curves of \cen in the 2--8 keV energy bands summed over three modules of \ixpe. Vertical dashed lines define good time intervals applied for the polarimetric analysis (see text for details).  Times of eclipses of XRP by the optical companion are marked with vertical shaded stripes. 
Observations performed in January and July 2022 are shown on the left and right panels, respectively. 
 }
 \label{fig:ixpe-lc}
\end{figure*}
%%%%%%%%%%%%%%%%%%%%%%%%%

Here we present the results of the analysis of Cen X-3 observations by \ixpe performed in two very different intensity states. 
First, we describe the observations and data reduction procedures in Sect.~\ref{sec:data}. 
The results are  presented in Sect.~\ref{sec:res}. 
We discuss possible sources of the observed polarization and the geometry of the pulsar in Sect.~\ref{sec:discussion}.  
Finally, we give the summary in Sect.~\ref{sec:sum}.

%\clearpage 

\section{Data} 
\label{sec:data}

\ixpe is a NASA mission in partnership with the Italian space agency. It was launched by a Falcon 9 rocket on 2021 December 9. It consists of three grazing incidence telescopes, each comprising an X-ray mirror assembly and a polarization-sensitive detector unit (DU) equipped with a gas-pixel detector \citep[GPD, ][]{2021AJ....162..208S,2021APh...13302628B}, to provide imaging polarimetry over a nominal 2--8 keV band with time resolution of the order of 10~$\mu$s. 
A detailed description of the instrument and its performance is given in \citet{Weisskopf2022}.

\ixpe observed \cen twice over the periods of 2022 Jan 29--31 and July 4--7 with a total effective exposure of $\simeq$68~ks and $\simeq$178~ks, respectively. 
The data have been processed with the {\sc ixpeobssim} package version 28.0.0 \citep{2022arXiv220306384B} using CalDB released on 2022 March 14. Before the scientific analysis, the position offset correction and energy calibration were applied.

Source photons were collected in a circular region with radius $R_{\rm src}$ of 70\arcsec\ centered on the \cen position. 
Background counts were extracted from an annulus with the inner and outer radii of 2$R_{\rm src}$  and 4$R_{\rm src}$, respectively. In the 2--8 keV band, the background comprises about 10\% and 0.5\% of the total count rate in the source region in the first and second observations, respectively.
The event arrival times were corrected to the Solar system barycenter using the standard {\tt barycorr} tool from the {\sc ftools} package  and accounting for the effects of binary motion using the orbital parameters from \citet{2010MNRAS.401.1532R}. For the second observation we found residual regular variations of the spin frequency over the observation even after correction, so the mid-eclipse epoch was adjusted to improve orbital solution as described in Appendix~\ref{app:1}. 

The flux (Stokes parameter $I$) energy spectra have been binned to have at least 30 counts per energy channel. 
The same energy binning was also applied to the spectra of the Stokes parameters $Q$ and $U$. Taking into account the high number of the source counts and low background level, the unweighted approach has been applied \citep{Di_Marco_2022}. All the spectra were fitted with the {\sc xspec} package \citep{Arn96} using the instrument response functions of version 10 and a $\chi^2$ statistic. 
The uncertainties are given at the 68.3\% confidence level unless stated otherwise.

%%%%%%%%%%%%%%%%%%%%%%%%%%
\begin{figure}
\centering
\includegraphics[width=0.95\linewidth]{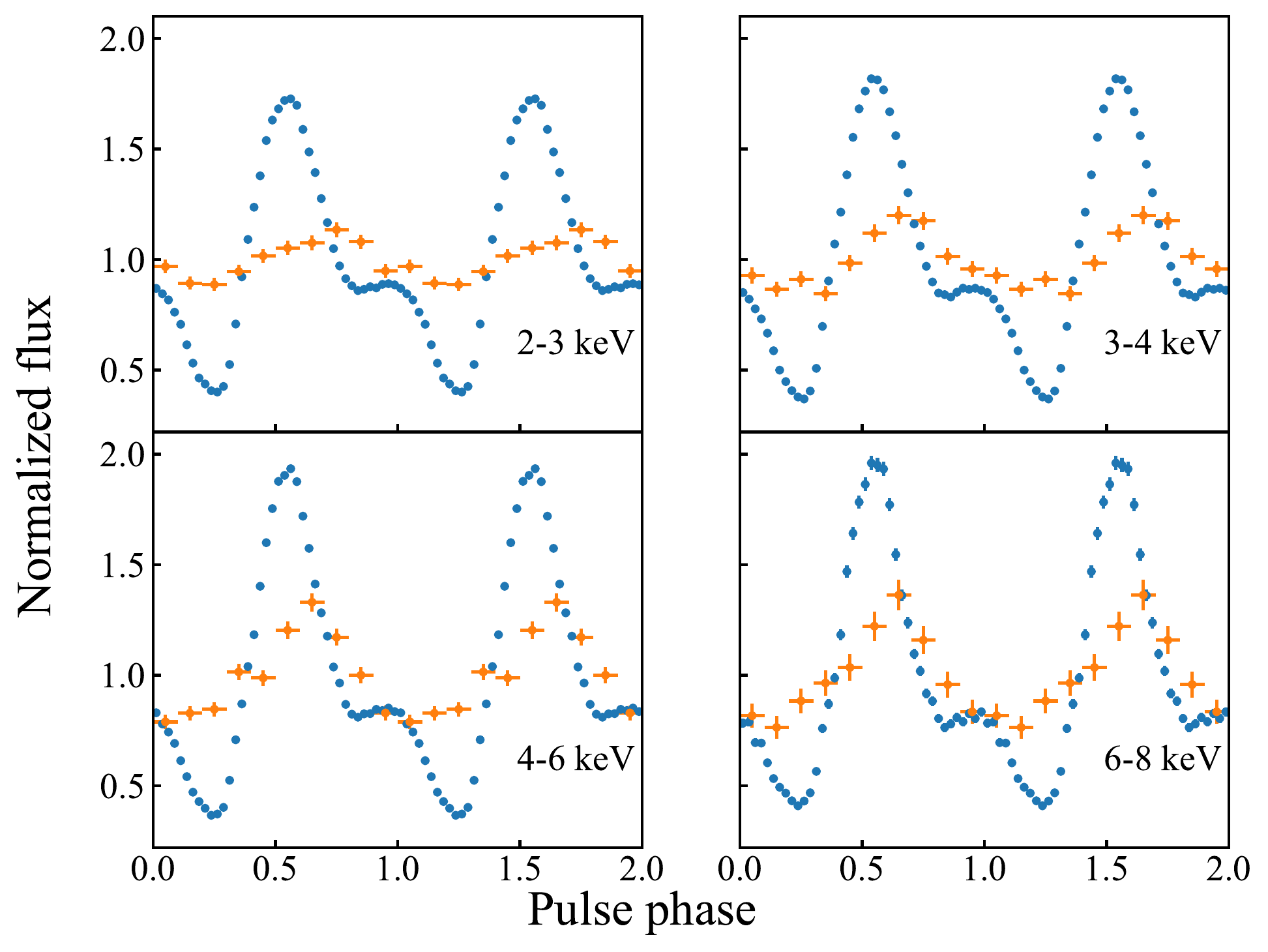}
\caption{Pulse profile of \cen in different energy bands as seen by \ixpe in the low (orange) and bright states (blue). Data from the three telescopes were combined. 
}
 \label{fig:ixpe-pprof}
\end{figure}
%%%%%%%%%%%%%%%%%%%%%%%%%

\section{Results} 
\label{sec:res}

The light curve of \cen obtained with the \ixpe observatory in the 2--8 keV band is shown in Figure~\ref{fig:ixpe-lc}.
The source was observed in January and July 2022 in two states different by a factor of $\sim30$ in mean off-eclipse count rate. During both observations a sharp drop by almost two orders of magnitude associated with the eclipse of the pulsar by the companion occurred several times. The time intervals affected by the eclipses ingress and egress were excluded from the following analysis (i.e. only data outside of the eclipse as marked with the dashed lines in Figure~\ref{fig:ixpe-lc} was used). To study the effect of the different mass accretion rate on the polarization properties, the two observations were analyzed independently.

In the low and bright states of the source we were able to measure the spin period of the pulsar with high accuracy, $P_{\rm spin-low}=4.79672(4)$~s and $P_{\rm spin-high}=4.7957473(8)$~s, respectively. 
In the bright state the spin period was found to decrease during the observation with $\dot{P}_{\rm spin-high}=-1.45(5)\times10^{-10}$\,s\,s$^{-1}$.
The pulsed fraction, defined as $PF = (F_{\max} - F_{\min})/(F_{\max} + F_{\min})$, where  $F_{\max}$ and $F_{\min}$ are the maximum and minimum count rates in the pulse profile, was found to be significantly different in the two observations: $18\pm2$\% in the low state and $66.5\pm0.6$\% in the bright state.  
The resulting pulse profiles in four energy bands for different luminosity states are shown in Figure~\ref{fig:ixpe-pprof}.

Polarimetric analysis was performed using two approaches implemented in the {\sc ixpeobssim} package: (i) a simplified approach 
based on the formalism of \cite{2015APh....68...45K} (the {\tt pcube} algorithm in the {\tt xpbin} tool); (ii) spectro-polarimetric analysis using {\sc xspec} \citep[see][]{2017ApJ...838...72S}, taking into account background and proper response functions of the instrument (the PHA1, PHA1Q, and PHA1U algorithms in the {\tt xpbin} tool). Below we present results obtained from both methods.

%%%%%%%%%%%%%%%%%%%%%%%%%%
\begin{figure}
\centering
\includegraphics[width=0.95\linewidth]{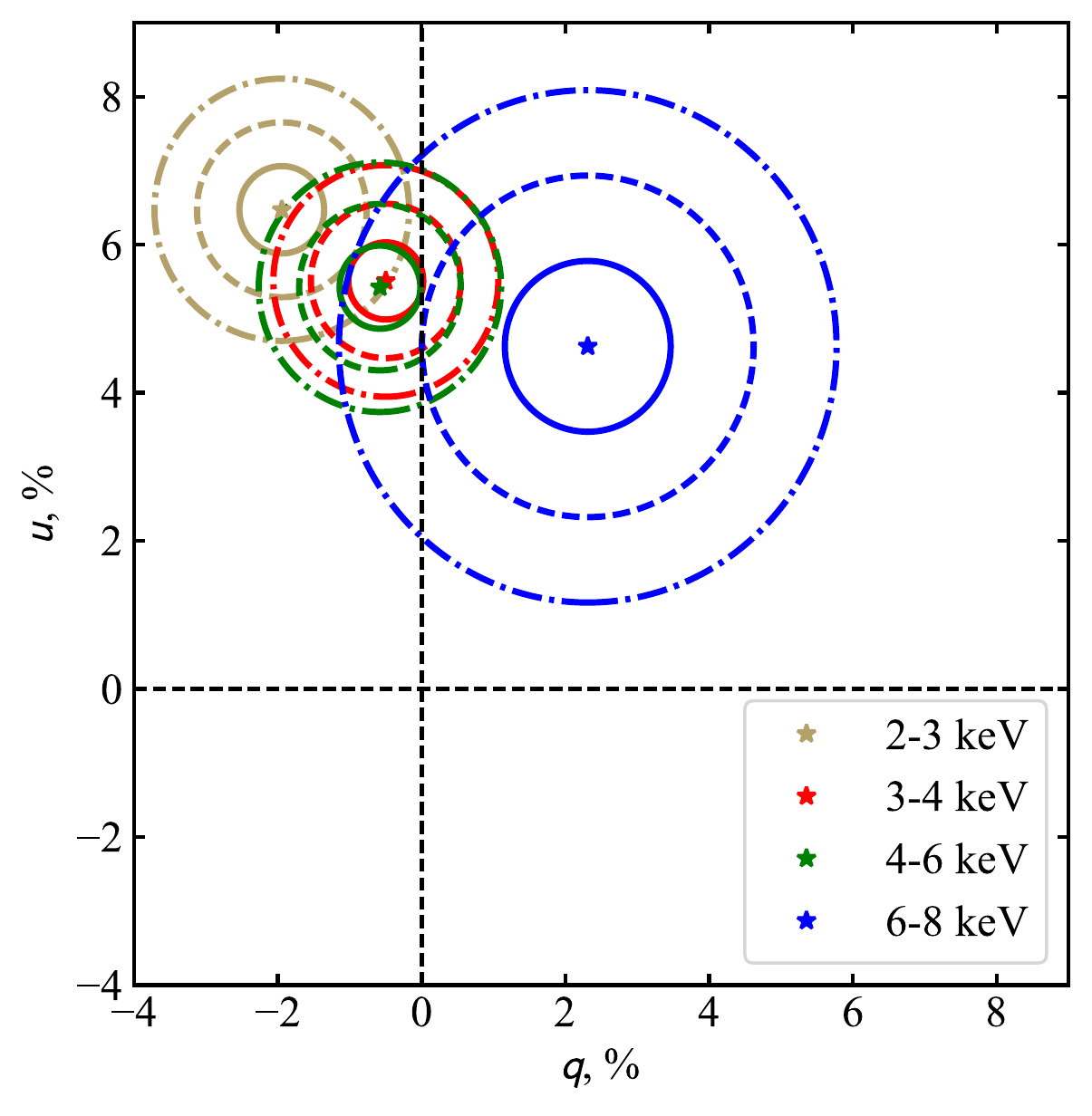}
\caption{Dependence of the phase-averaged normalized Stokes parameters $q$ and $u$ on energy in the bright state of \cen. Solid, dashed and dashed-dotted lines correspond to the 1, 2 and 3$\sigma$ confidence levels, respectively.
Data from the three telescope units were summed together. 
 }
 \label{fig:ixpe-polar}
\end{figure}
%%%%%%%%%%%%%%%%%%%%%%%%%

%%%%%%%%%%%%%%%%%%%%%%%%
\begin{table}
\centering  
\caption{Measurements of the PD, PA and MDP$_{99}$ in different  states of \cen as a function of energy.  }
\label{tab:enres}   
\begin{tabular}{cccc}
\hline \hline
Energy  & PD & PA & MDP$_{99}$ \\
 (keV) &  (\%) & (deg) &  (\%) \\
\hline
\multicolumn{4}{c}{Low state (January 2022)}  \\
%\hline
2--3 & $6$ & - & $18$ \\
3--4 & $4$ & - & $15$ \\ 
4--6 & $9$ & - & $12$ \\
6--8 & $12$ & - & $18$ \\
\hline
\multicolumn{4}{c}{Bright state (July 2022)}  \\
%\hline
2--3 & $6.8\pm0.6$ & $53.4\pm2.5$ & 1.8 \\
3--4 & $5.5\pm0.5$ & $47.6\pm2.7$ & 1.6 \\ 
4--6 & $5.5\pm0.6$ & $48.0\pm3.0$ & 1.7 \\
6--8 & $5.2\pm1.2$ & $31.7\pm6.4$ & 3.5 \\
\hline 
\end{tabular}
\end{table}
%%%%%%%%%%%%%%%%%%%%%%%%

\subsection{Simplified polarimetric analysis}
\label{sec:pcube}

%%%%%%%%%%%%%%%%%%%%%%%%%%
\begin{figure}
\centering
\includegraphics[width=0.95\linewidth]{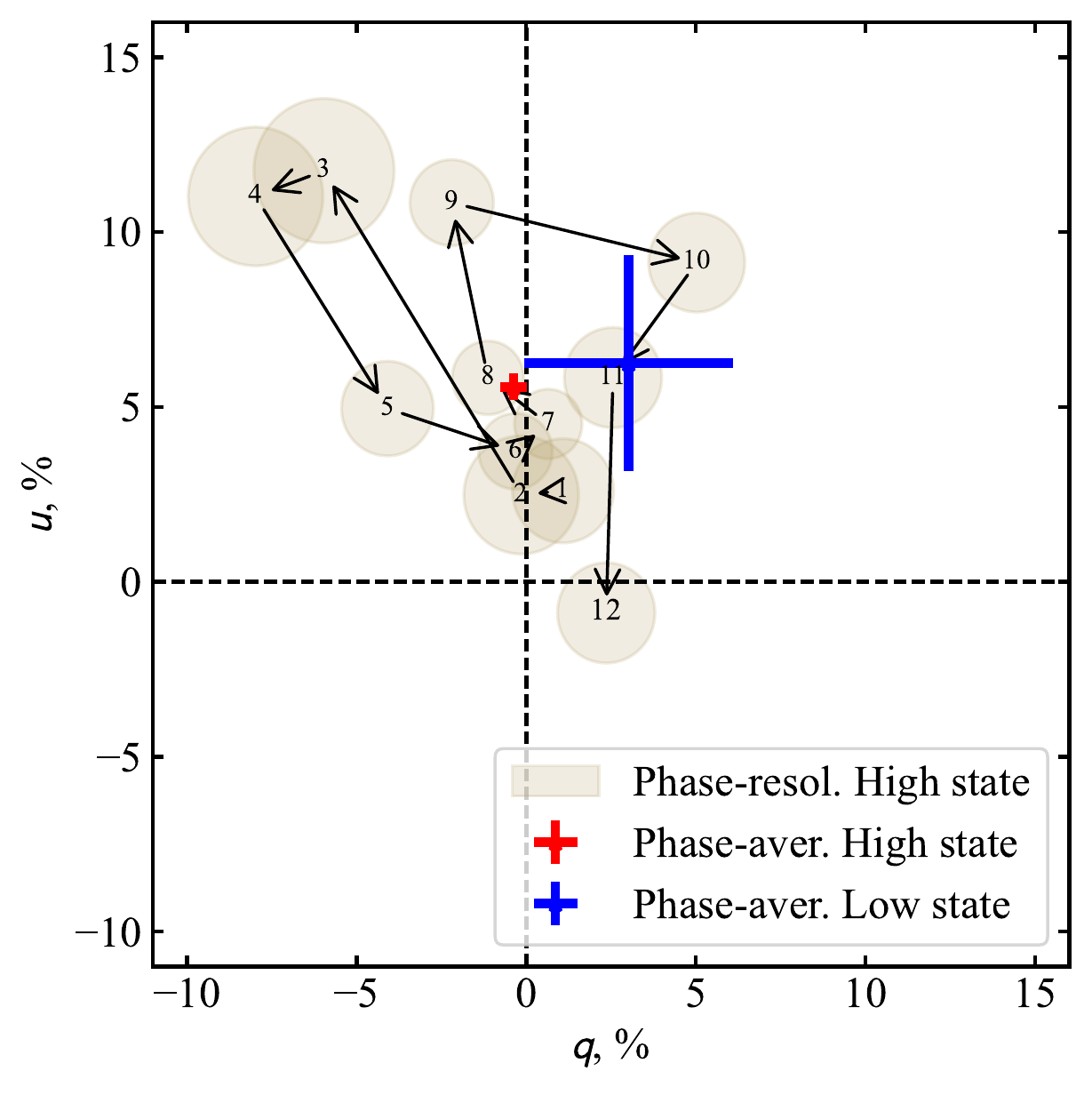}
\caption{Variations of the phase-resolved normalized Stokes parameters $q$ and $u$ with pulsar phase in the 2--8 keV energy band averaged over all DUs (gray circles and arrows) measured in the bright state of \cen.
The phase-averaged values for the bright and low states are shown in red and blue color, respectively.  
 }
 \label{fig:pcub-aver}
\end{figure}
%%%%%%%%%%%%%%%%%%%%%%%%%

To study the polarimetric properties of \cen we started with the standard  analysis using the formalism of \cite{2015APh....68...45K}. 
First we binned the data collected in different intensity states into four energy bins averaging over the spin phase. The resulting phase-averaged energy dependence of the normalized Stokes parameters $q=Q/I$ and $u=U/I$ measured in the bright state is shown in Figure~\ref{fig:ixpe-polar}. 
The corresponding values of PD and polarization angles (PA, measured from north to east) are presented in Table~\ref{tab:enres}. In the low state the estimated PD is not significant and below the minimum detectable polarization at the 99\% confidence level (MDP$_{99}$) of $\sim$15--20\% in each energy band. In this case we do not present the corresponding PA as it has no physical meaning when the polarization signal is not significantly detected. In the bright state both PD and PA are well constrained in all considered energy bands.

%%%%%%%%%%%%%%%%%%%%%%%%%%
\begin{figure}
\centering
\includegraphics[width=0.95\linewidth]{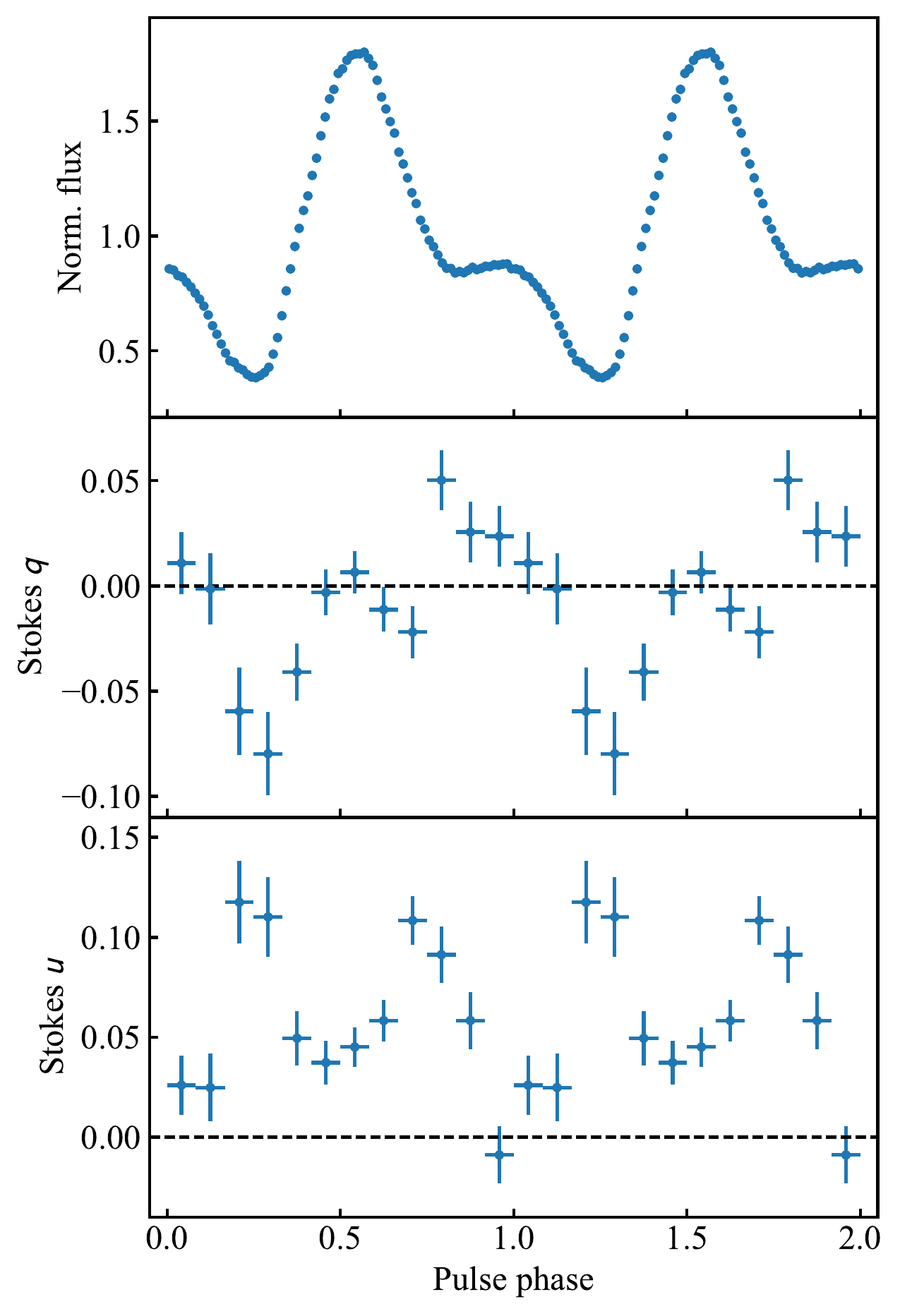}
\caption{Dependence of the normalized flux, normalized Stokes parameters $q$ and $u$ on the pulse phase in the 2--8 keV energy band in the bright state of \cen estimated using the formalism of \cite{2015APh....68...45K}. 
Data from the three \ixpe telescopes are combined. 
 } 
 \label{fig:ixpe-st.pd.pa}
\end{figure}
%%%%%%%%%%%%%%%%%%%%%%%%%

By considering the full \ixpe energy range, 2--8 keV, we again did not find a significant polarization signal in the low state  with a corresponding formal value of PD=$7\pm3$\%.  
%and MDP$_{99}$=9\%. 
However, in the bright state polarization was significantly detected at the $15\sigma$ level with PD=$5.6\pm0.4$\% and PA=$47.0\pm2.0$ deg. The results for both observations in the $q$-$u$ plane are shown in Figure~\ref{fig:pcub-aver}.
Because there is a detection of the polarization signal only in the data collected in July (i.e. the bright state),  we present below the results corresponding to this part of the dataset, if not stated otherwise. 

Given the strong angular dependence of the scattering cross-sections, the polarization properties of XRPs are expected to vary with the pulse phase. Therefore, as the next step we performed the phase-resolved polarimetric analysis using the same {\tt pcube} algorithm. In particular, using $P_{\rm spin-high}$, we calculated the pulse phase for each event and binned the data into 12 phase bins in the 2--8 keV energy band. The results of this analysis are shown in Figures~\ref{fig:pcub-aver} and \ref{fig:ixpe-st.pd.pa}. One can see that the normalized $q$ and $u$ Stokes parameters are strongly variable over the pulse phase.

%%%%%%%%%%%%%%%%%%%%%%%%%%
\begin{figure*}
\centering
\includegraphics[width=0.32\linewidth]{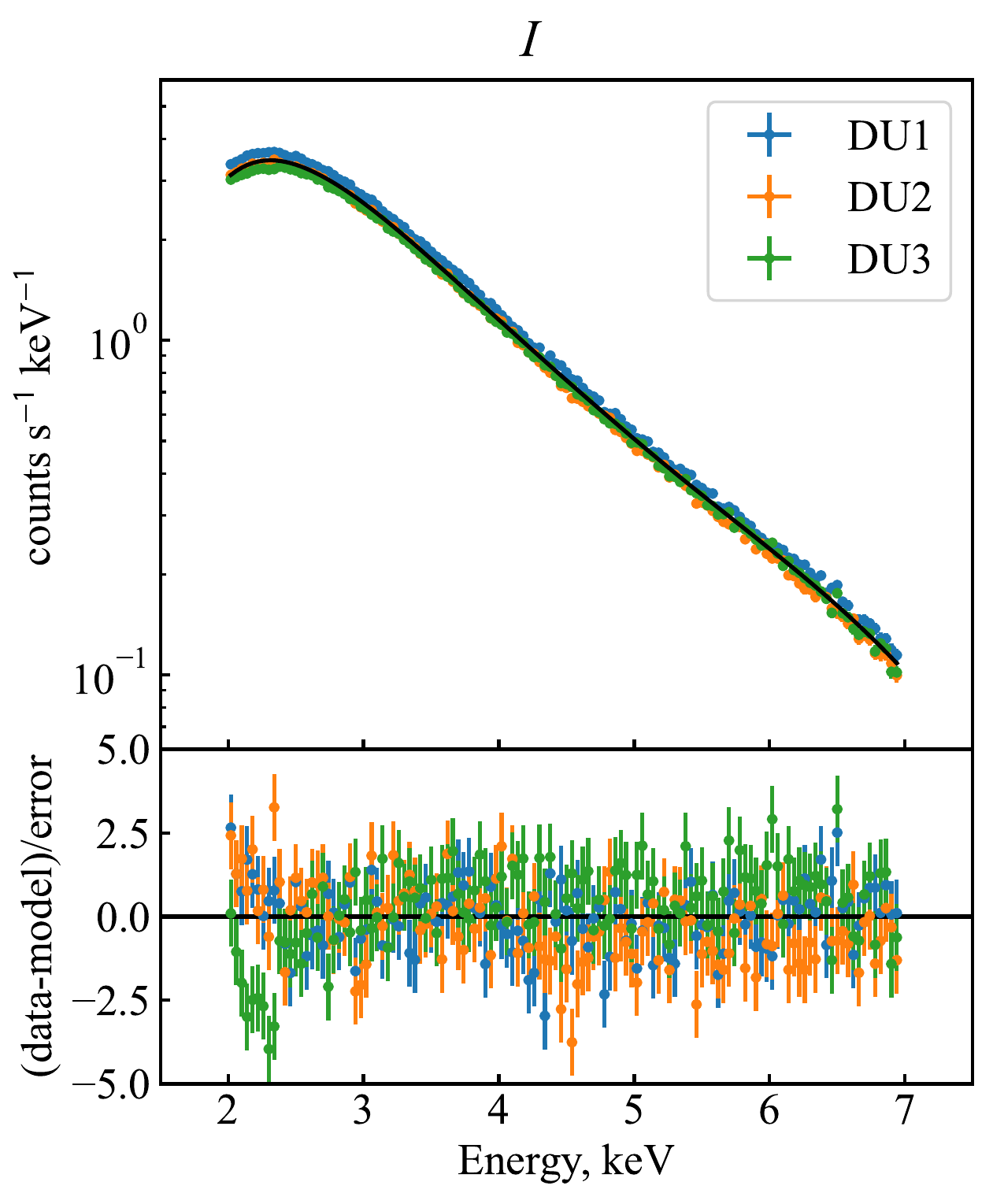}
\includegraphics[width=0.32\linewidth]{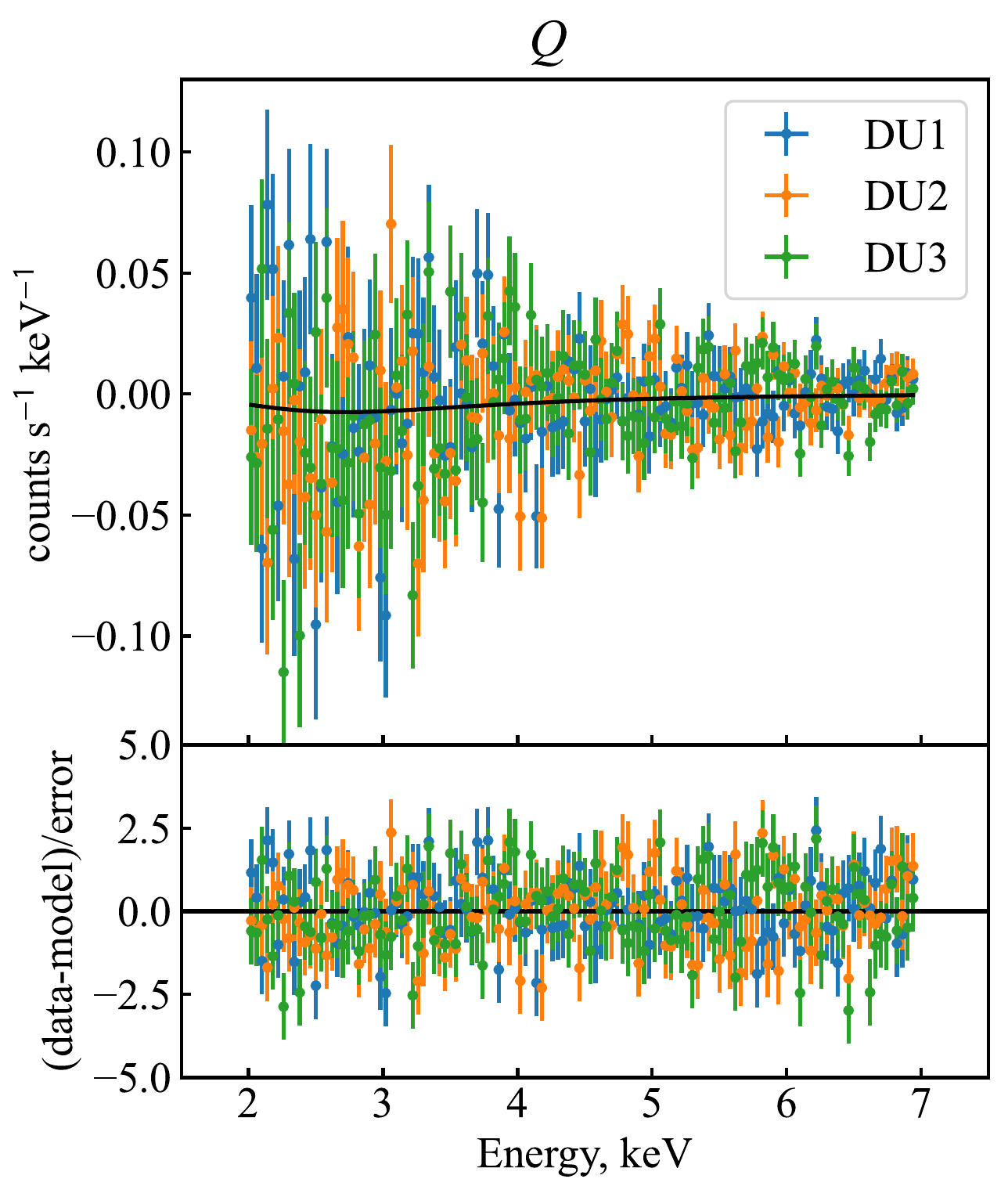}
\includegraphics[width=0.32\linewidth]{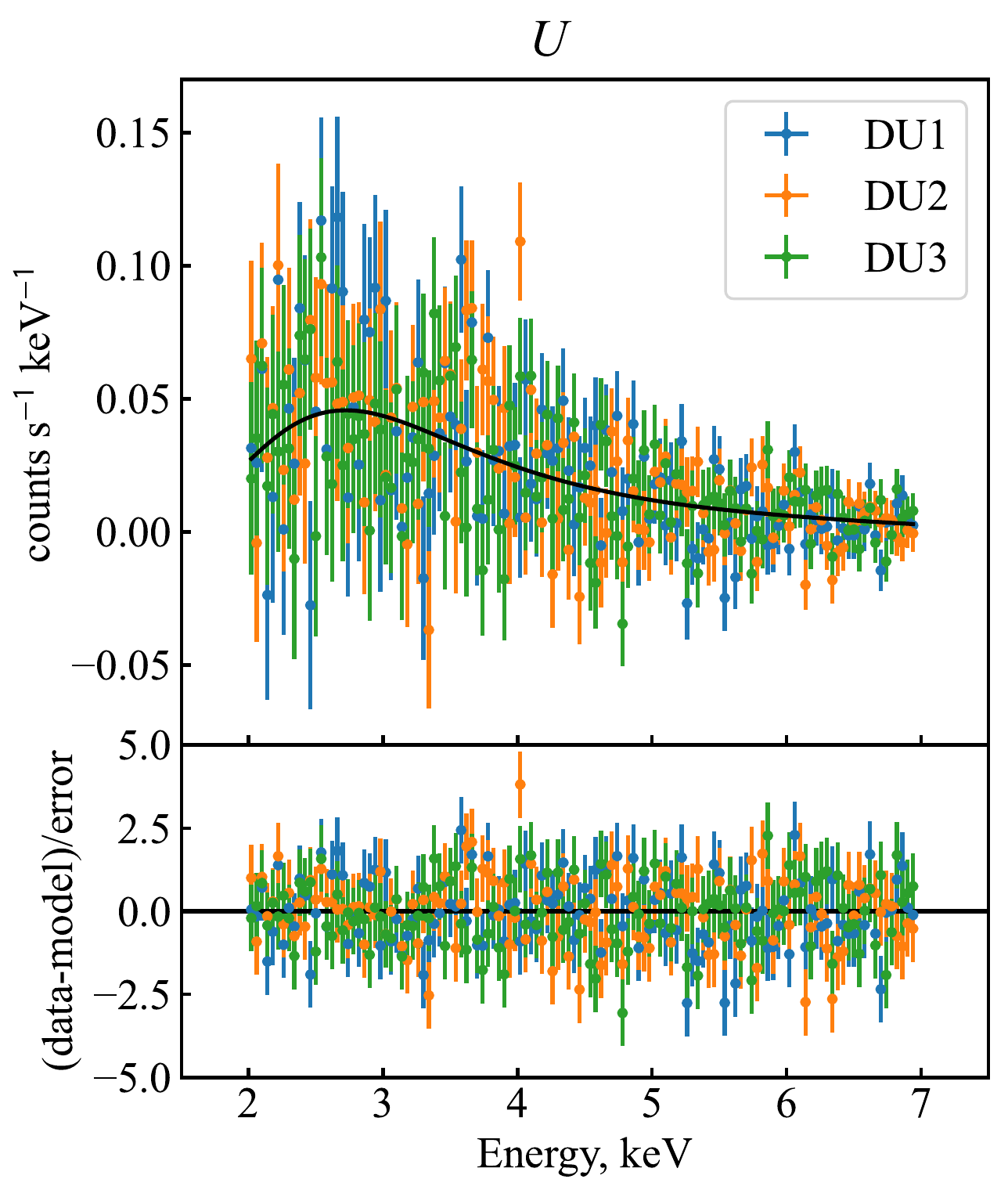}
\caption{Energy distributions of the Stokes parameters $I$, $Q$ and $U$ for the bright state of \cen superimposed to the best-fit model (top panels). The residuals between data and model normalized for the errors are shown in the bottom panels. The different colors represent the three \ixpe detectors: DU1 in blue, DU2 in orange and DU3 in green.}
 \label{fig:spec-aver}
\end{figure*}
%%%%%%%%%%%%%%%%%%%%%%%%%

%%%%%%%%%%%%%%%%%%%%%%%%%%
\begin{table}
    \caption{Spectral parameters for the best-fit model  obtained from {\sc xspec} for two intensity states of the source; uncertainties are at 68.3\% CL.}
    \label{tab:spec-aver}
    \centering
    \begin{tabular}{rll}
    \hline\hline
    Parameter & Value & Units\\ \hline
    \multicolumn{3}{c}{Low state (January 2022)}  \\
 %   \hline
    $N_{\rm H}$  & 0.6$^{+0.6}_{-0.4}$ &  $10^{22}$~cm$^{-2}$ \\
    const$_{\rm DU2}$ & 0.99$\pm0.02$ & \\
    const$_{\rm DU3}$ & 0.91$\pm0.01$ & \\
    Photon index & $-0.34\pm0.03$ & \\
    Fe line $E$  & 6.22$\pm0.08$ & keV \\ 
    Fe line $\sigma$  & 0.4$^{+0.1}_{-0.2}$ & keV \\ 
    Fe line norm  & 1.7$\pm0.3$ & $10^{-3}$ ph~cm$^{-2}$~s$^{-1}$\\ 
    PD & 3.9$\pm2.8$ & \% \\
    PA & unconstrained & deg \\
    Flux (2--8~keV) & 2.17$\pm0.03$& $10^{-10}$ erg~cm$^{-2}$~s$^{-1}$ \\
    Luminosity (2--8~keV) & $1.1\times10^{36}$  & erg~s$^{-1}$ at $d=6.4$~kpc\\ 
    $\chi^{2}$ (d.o.f.) & 1079 (1089) & \\
    \hline
    \multicolumn{3}{c}{Bright state (July 2022)}  \\
%    \hline
    $N_{\rm H}$  & $2.85\pm0.03$ &  $10^{22}$~cm$^{-2}$ \\
    const$_{\rm DU2}$ & 0.963$\pm0.002$ & \\ 
    const$_{\rm DU3}$ & 0.909$\pm0.002$ & \\ 
    Photon index & $1.32\pm0.01$ & \\ 
    PD & 5.8$\pm0.3$ & \% \\ 
    PA & 49.6$\pm1.5$ & deg \\ 
    Flux  (2--8~keV) & 38.56$\pm0.06$ & $10^{-10}$ erg~cm$^{-2}$~s$^{-1}$ \\ 
    Luminosity (2--8~keV) & $1.9\times10^{37}$  & erg~s$^{-1}$ at $d=6.4$~kpc\\ 
    $\chi^{2}$ (d.o.f.) & 1275 (1109) & \\ 
    \hline 
    \end{tabular}
\end{table}
%%%%%%%%%%%%%%%%%%%%%%%%%

\subsection{Spectro-polarimetric analysis}

To perform the spectro-polarimetric analysis, the source and background energy spectra were extracted for each DU using 
the {\tt PHA1}, {\tt PHA1Q}, and {\tt PHA1U} algorithms in the {\tt xpbin} tool and fitted simultaneously in {\sc xspec}. 
We restricted our spectral analysis to the 2--7\,keV energy band, ignoring photons at higher energies due to remaining calibration uncertainties. 
The energy spectrum of \cen is known to have a complicated structure consisting of several components \citep[see, e.g.,][]{Sanjurjo-Ferrin21}. However, the limited energy range of \textit{IXPE} and relatively low counting statistics allow us to use a much simpler model consisting of a power law modified by the interstellar absorption \citep[{\tt tbabs} in {\sc xspec} with abundances adopted from][]{Wilms2000} and an iron emission line in the form of a Gaussian. Due to the limited energy resolution of the instrument (FWHM $\sim15$\%  at 6 keV), the complex of iron lines  was fitted with a single broad component. 
The power-law component of the model was combined with a constant polarization model (energy-independent PD and PA), {\tt polconst} in {\sc xspec}. To account for possible discrepancies in calibration of independent DUs, a renormalization constant {\tt const}, different for each DU, was introduced as well; the constant was fixed to unity for DU1, taken as a reference. The final form of the model in {\sc xspec} is {\tt const$\times$tbabs$\times$(polconst$\times$powerlaw+gau}).

%%%%%%%%%%%%%%%%%%%%%%%%%%
\begin{figure}
\centering
\includegraphics[width=0.95\linewidth]{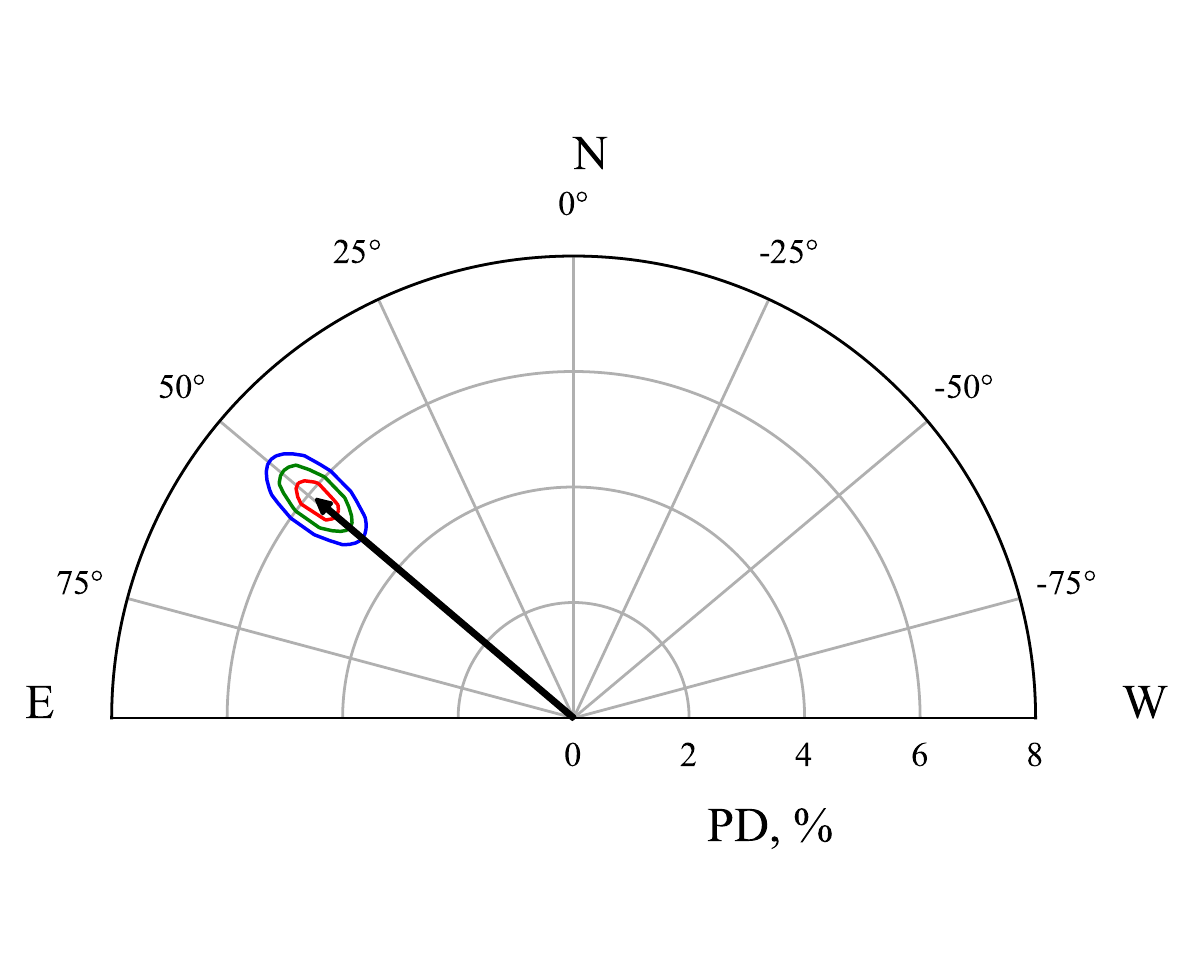}
\caption{Polarization vector of \cen based on the spectral fitting of the pulse phase-averaged data collected in the bright state. The PD and PA confidence levels contours at 1, 2 and 3$\sigma$ are presented in polar coordinates in red, green and blue color, respectively. }
 \label{fig:cont-aver}
\end{figure}
%%%%%%%%%%%%%%%%%%%%%%%%%

%%%%%%%%%%%%%%%%%%%%%%%%%%
\begin{figure*}%[b]
\centering
\includegraphics[width=0.3\linewidth]{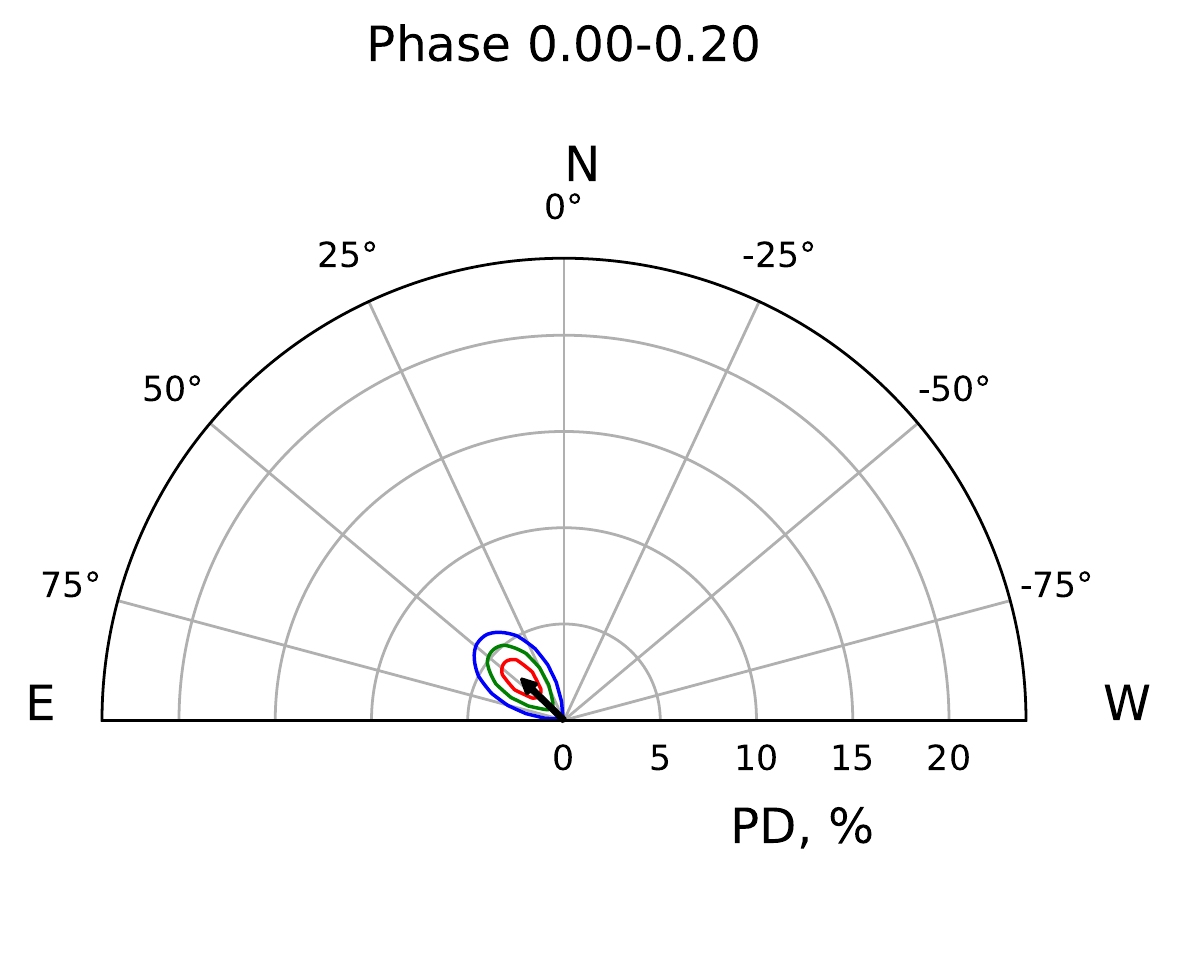}
\includegraphics[width=0.3\linewidth]{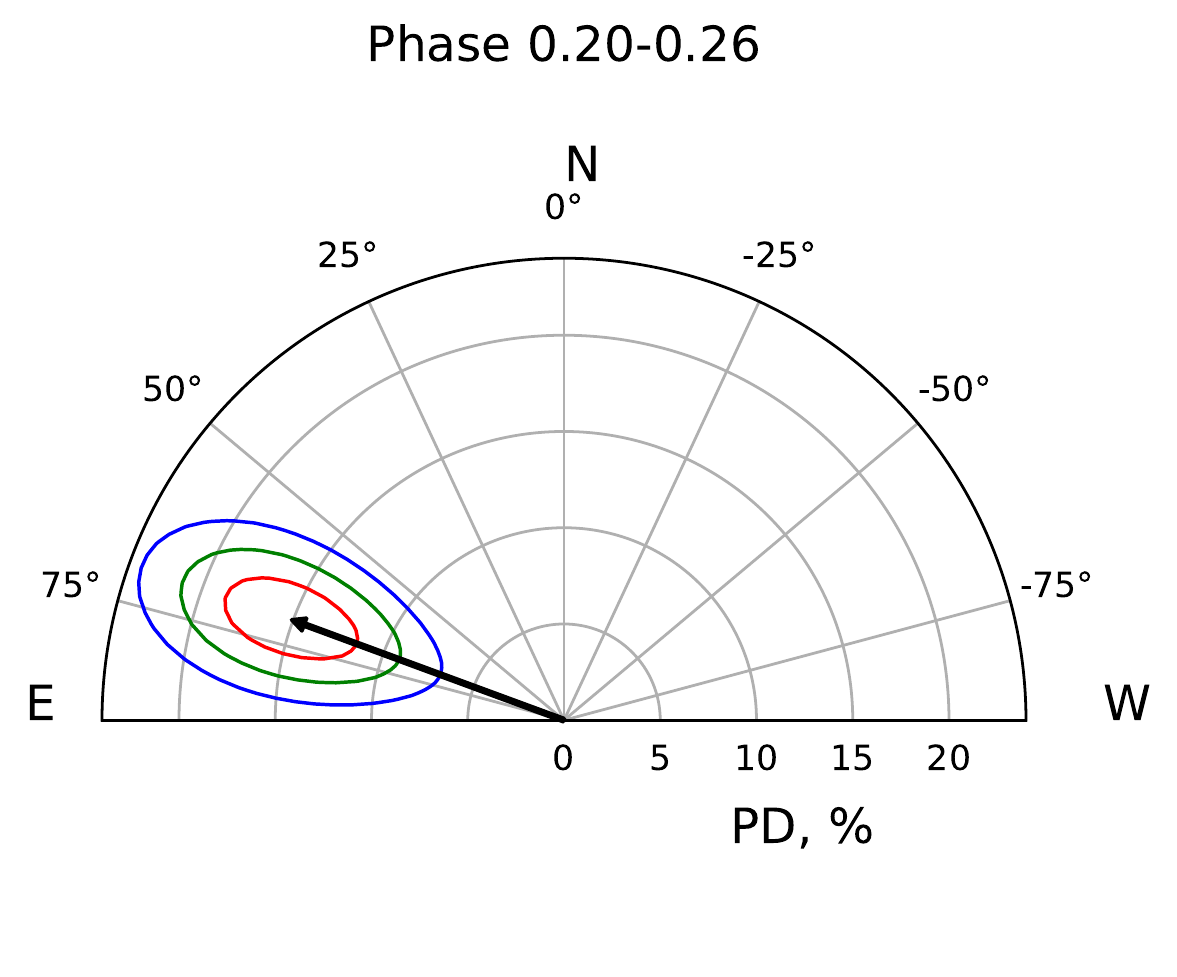}
\includegraphics[width=0.3\linewidth]{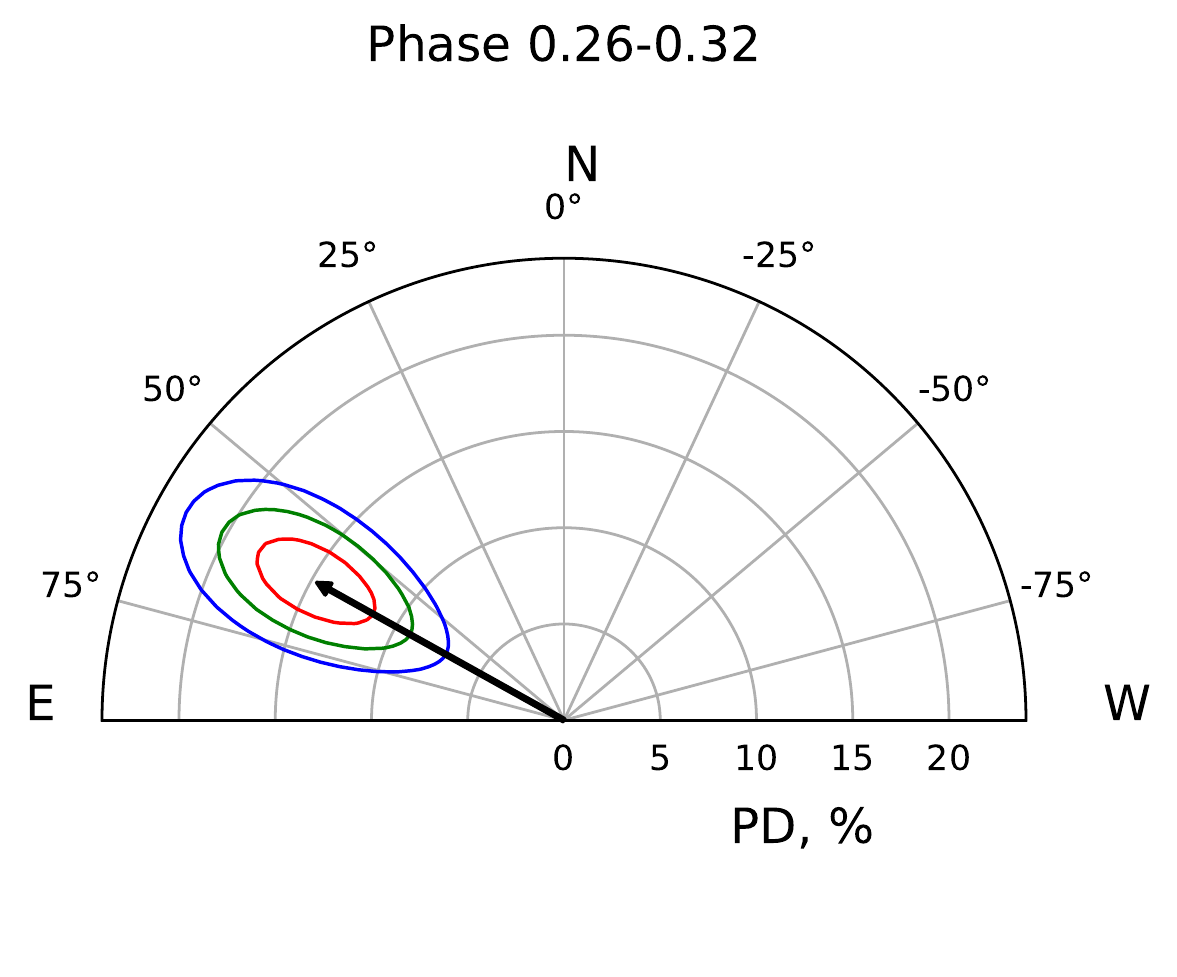}
\includegraphics[width=0.3\linewidth]{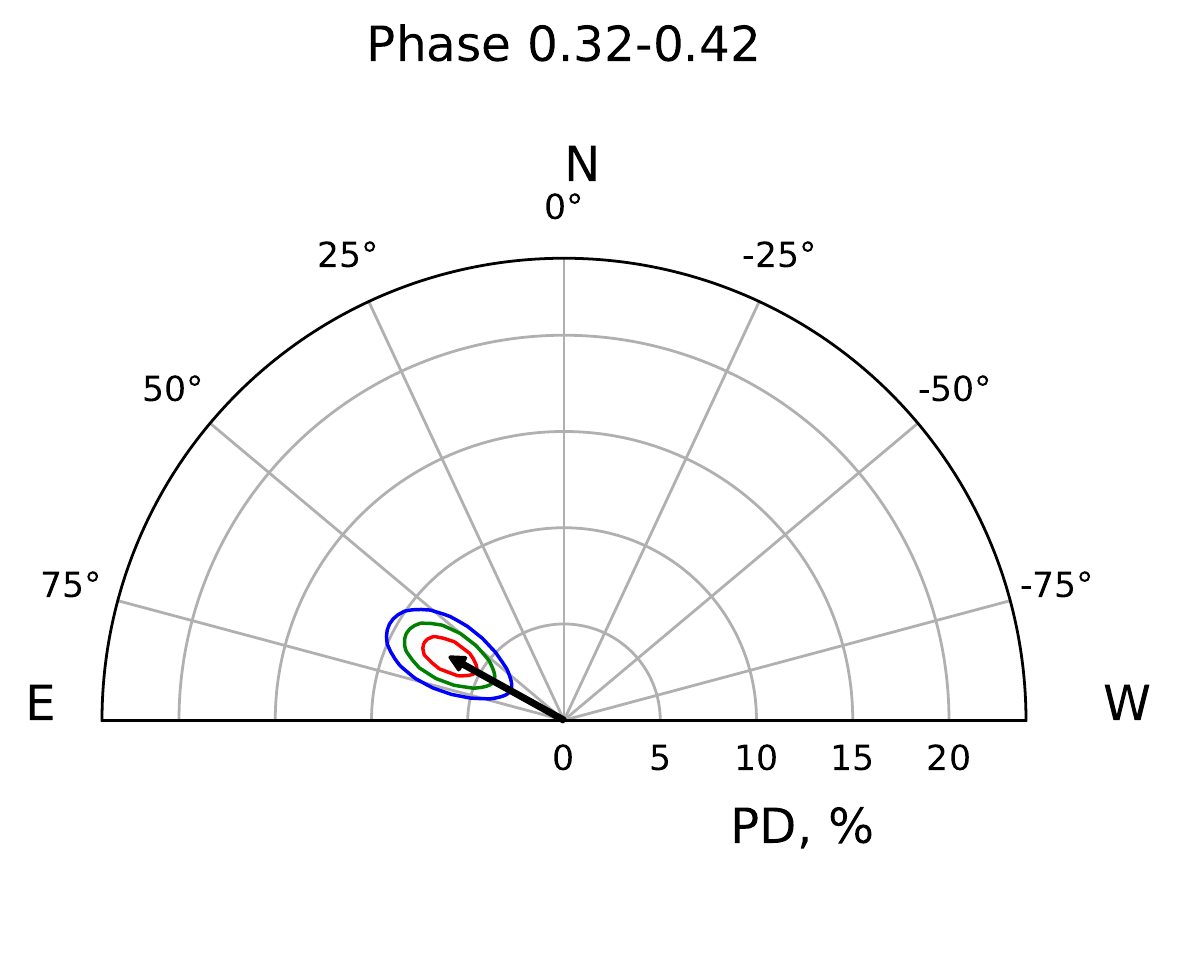}
\includegraphics[width=0.3\linewidth]{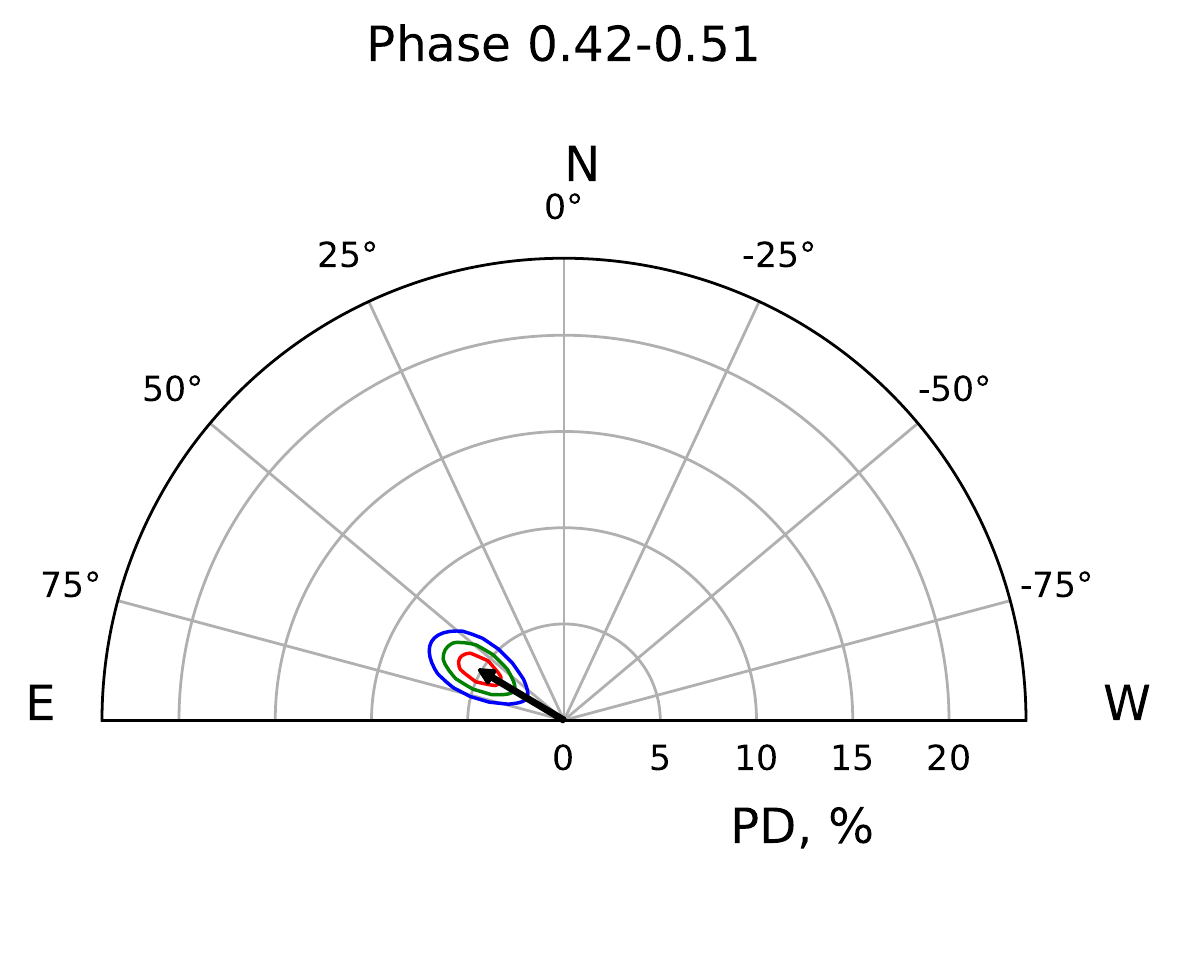}
\includegraphics[width=0.3\linewidth]{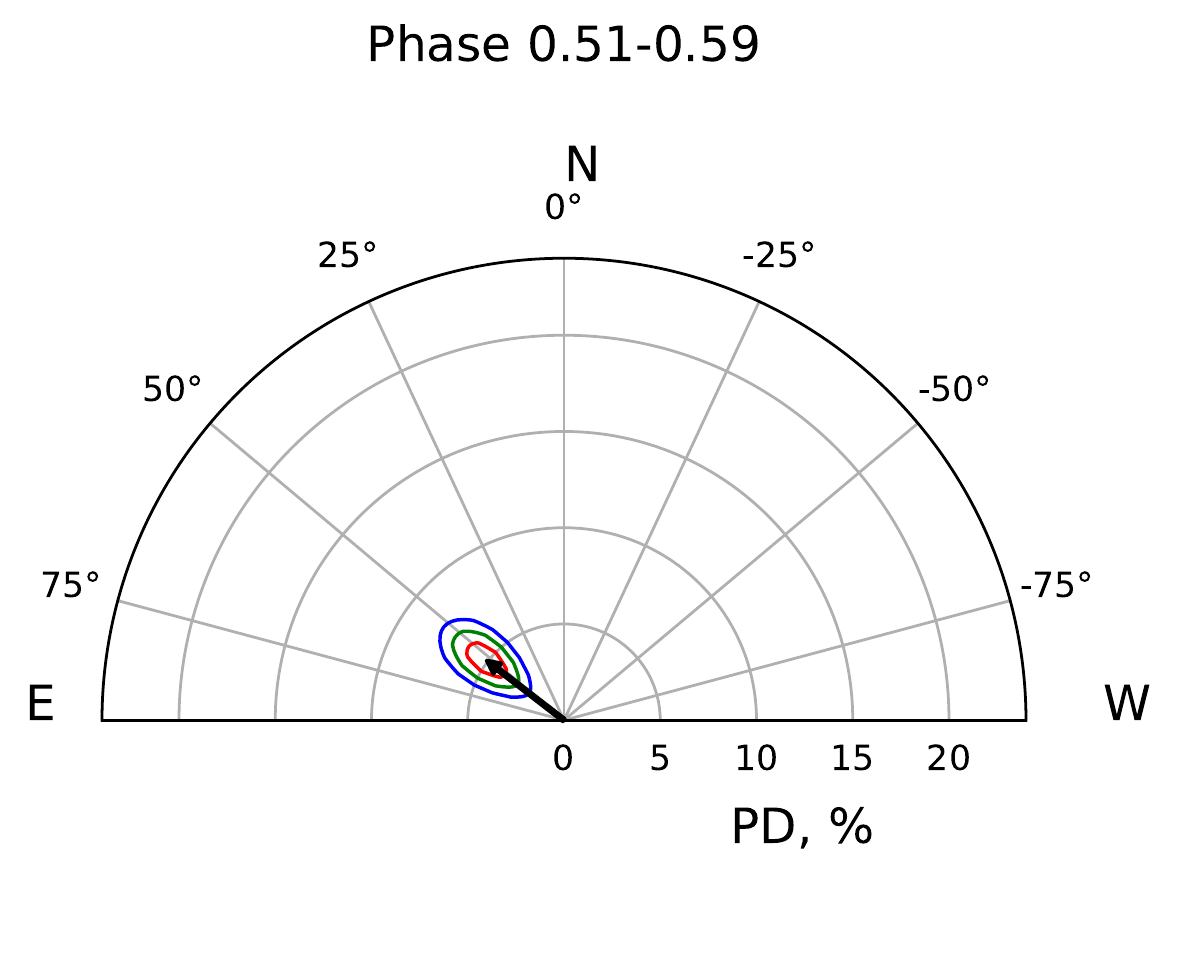}
\includegraphics[width=0.3\linewidth]{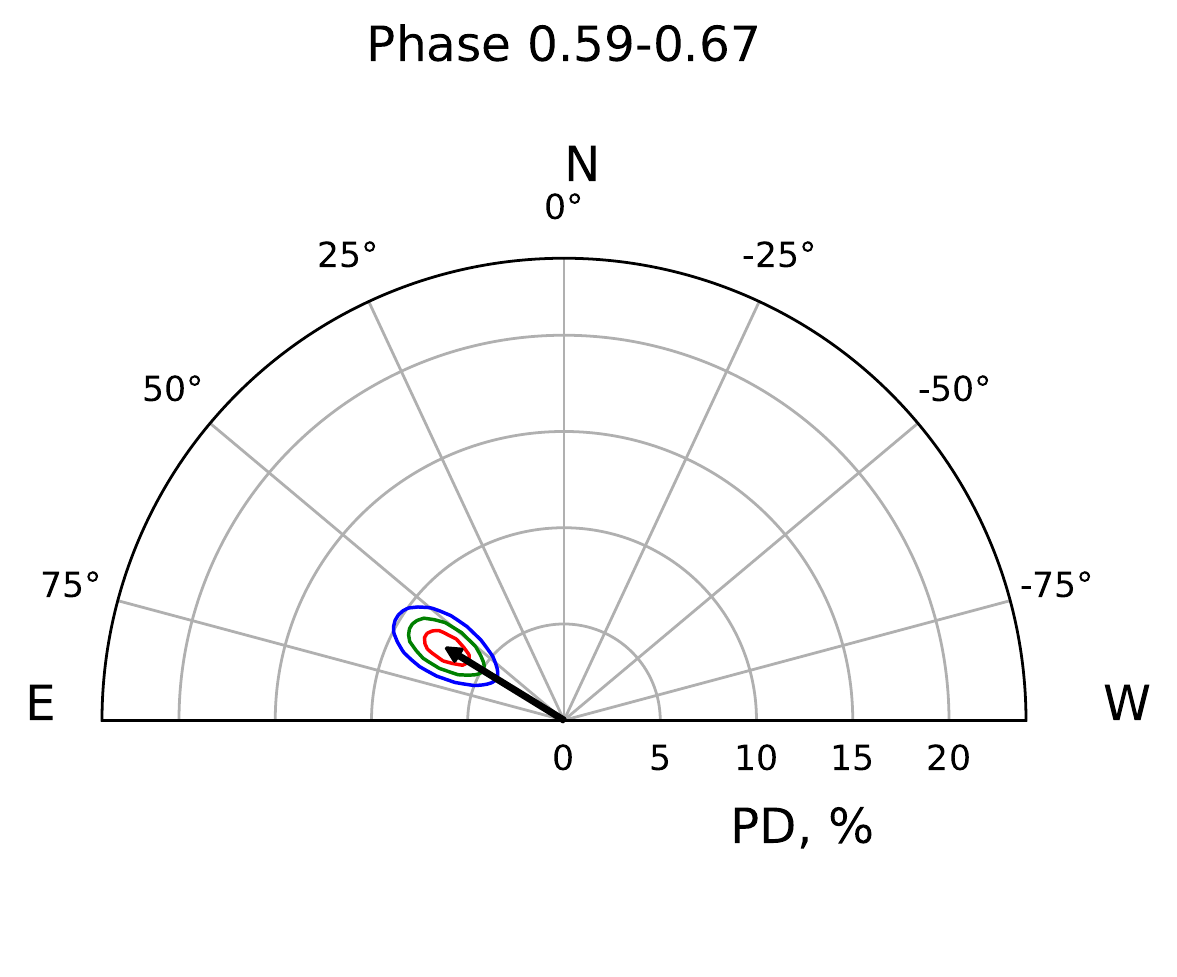}
\includegraphics[width=0.3\linewidth]{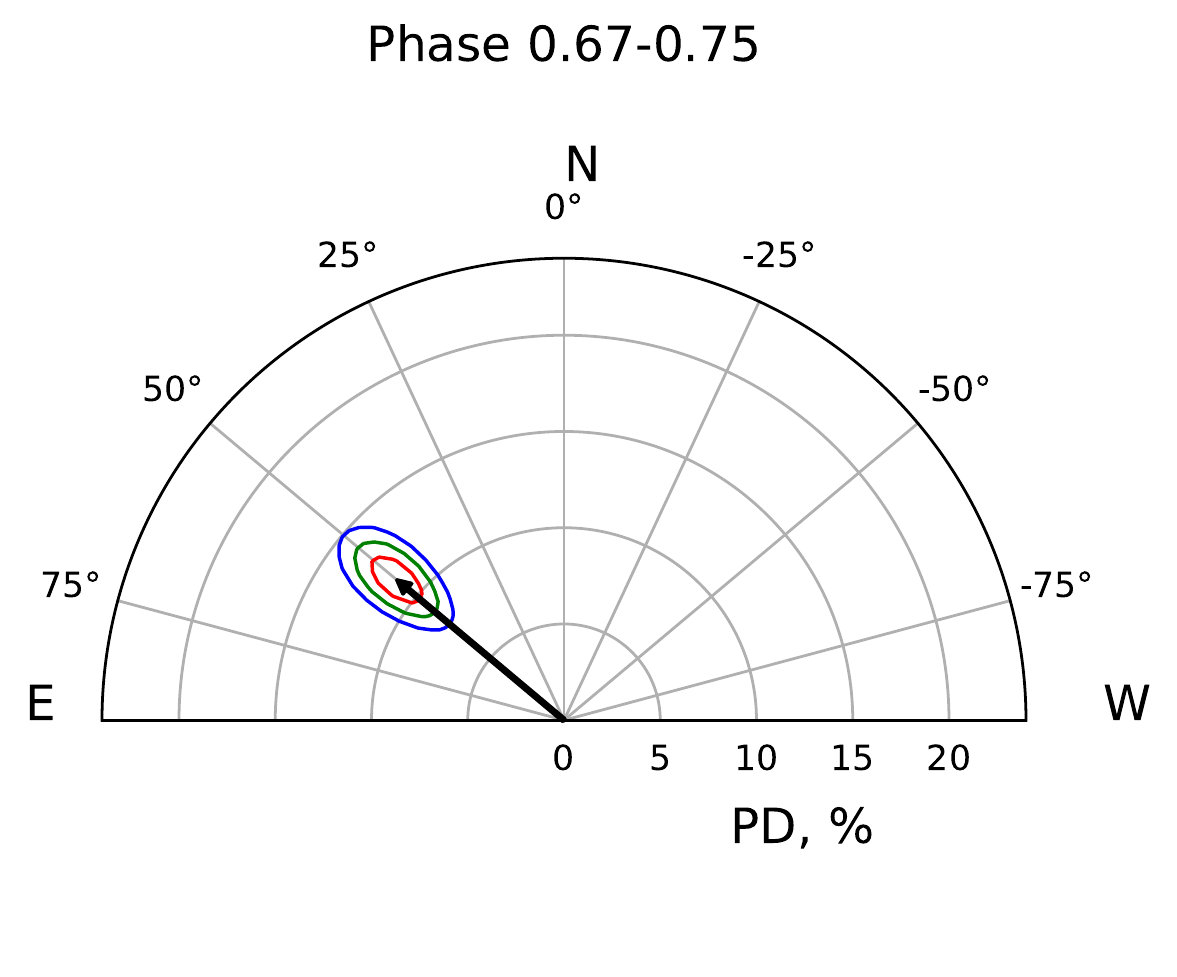}
\includegraphics[width=0.3\linewidth]{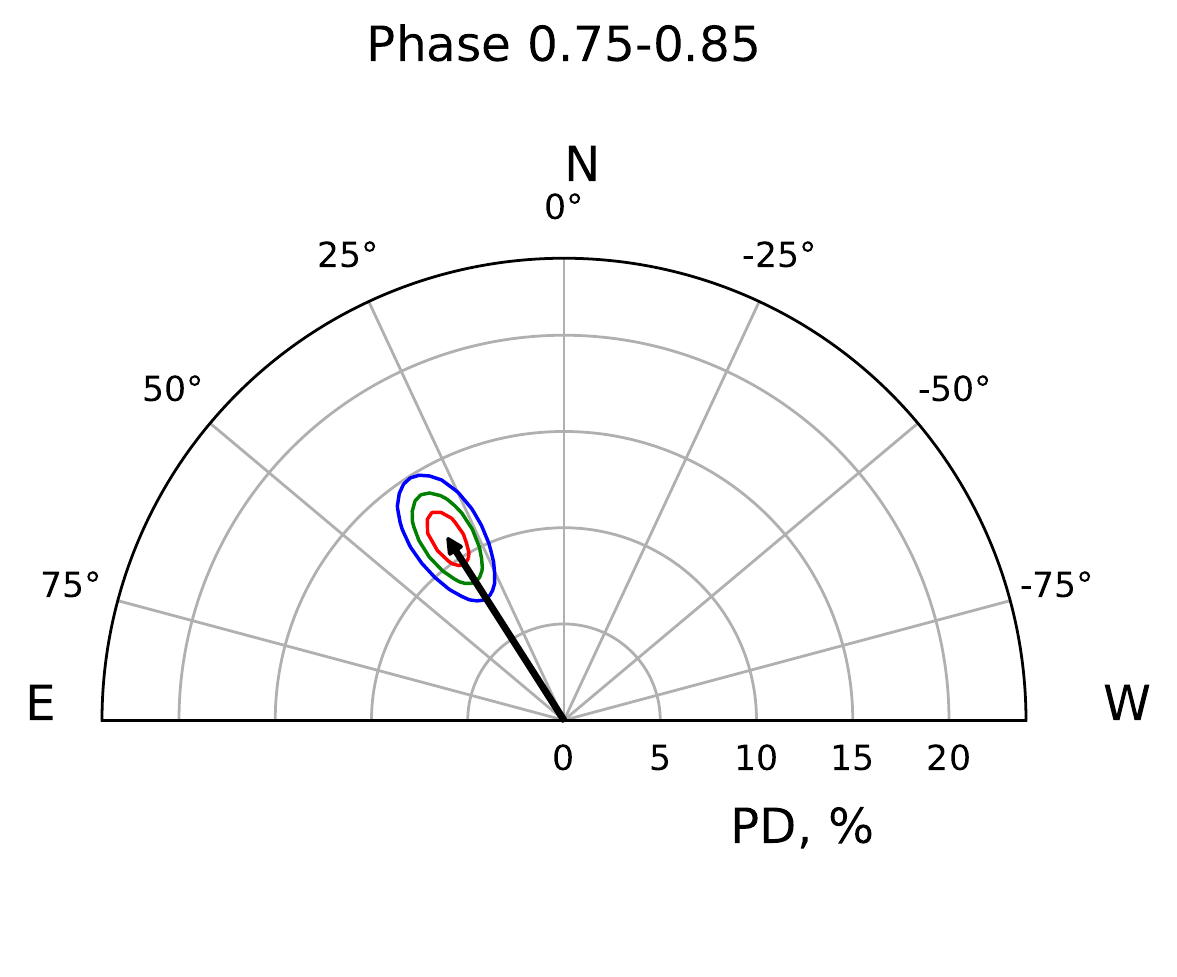}
\includegraphics[width=0.3\linewidth]{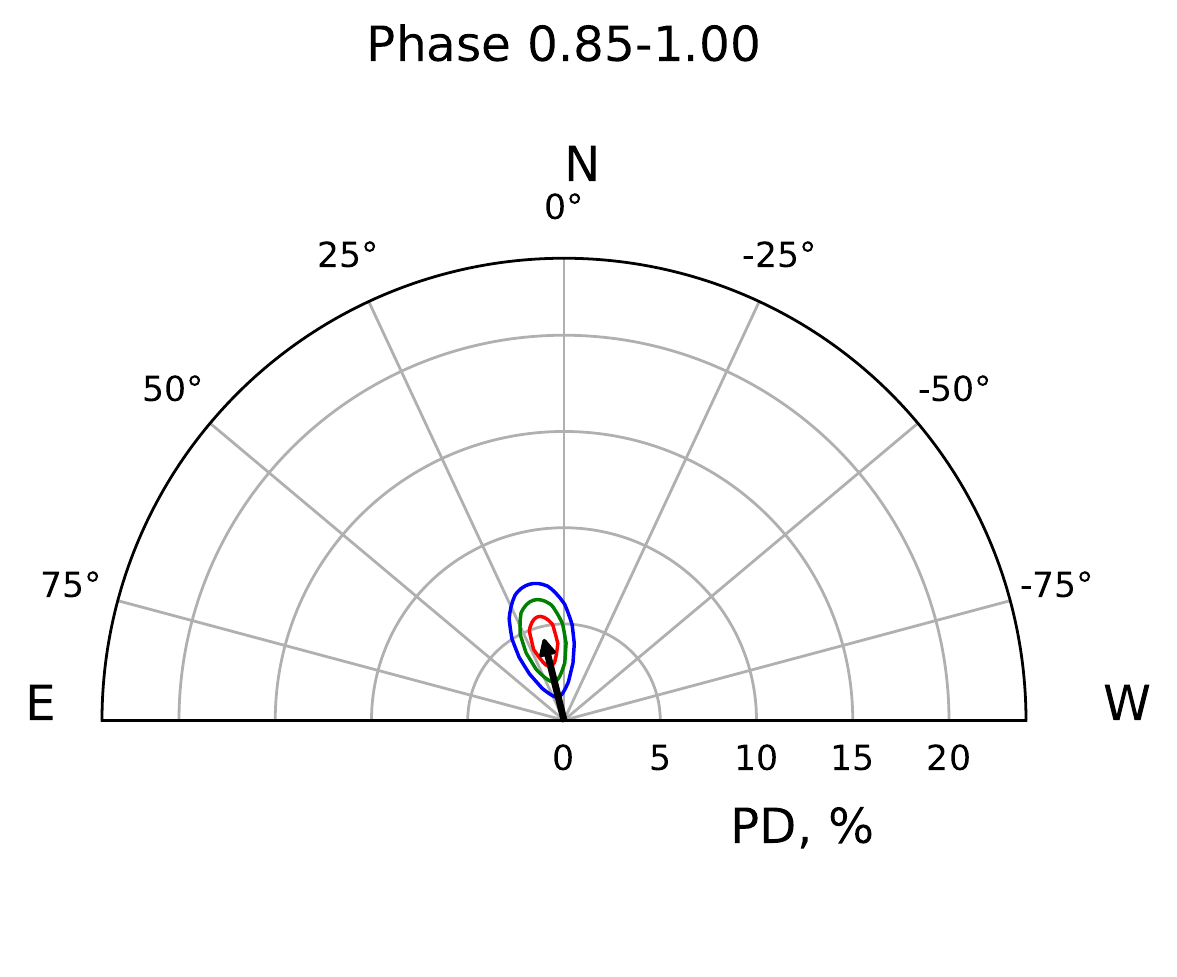}
\caption{Polarization vectors of \cen based on the spectral fitting of the pulse phase-resolved data in the bright state as a function of the phase.  In each plot the PD and PA confidence level contours at 1, 2 and 3$\sigma$ (in red, green and blue color, respectively) are shown in polar coordinates for 10 different phase intervals. 
}
 \label{fig:cont-resol}
\end{figure*}
%%%%%%%%%%%%%%%%%%%%%%%%%

First, we applied this model to the phase-averaged data. Likewise in the energy-binned analysis from the {\tt pcube} algorithm we investigated the properties of \cen in two intensity states independently. The parameters of the best-fit model are presented in Table~\ref{tab:spec-aver}. Similarly to the simplified polarimetric analysis, spectro-polarimetric analysis did not reveal significant polarization in the low state with the $3\sigma$ upper limit of 12\% and showed highly significant polarization in the bright state.

%%%%%%%%%%%%%%%%%%%%%%%%%%
\begin{figure}
\centering
\includegraphics[width=0.9\linewidth]{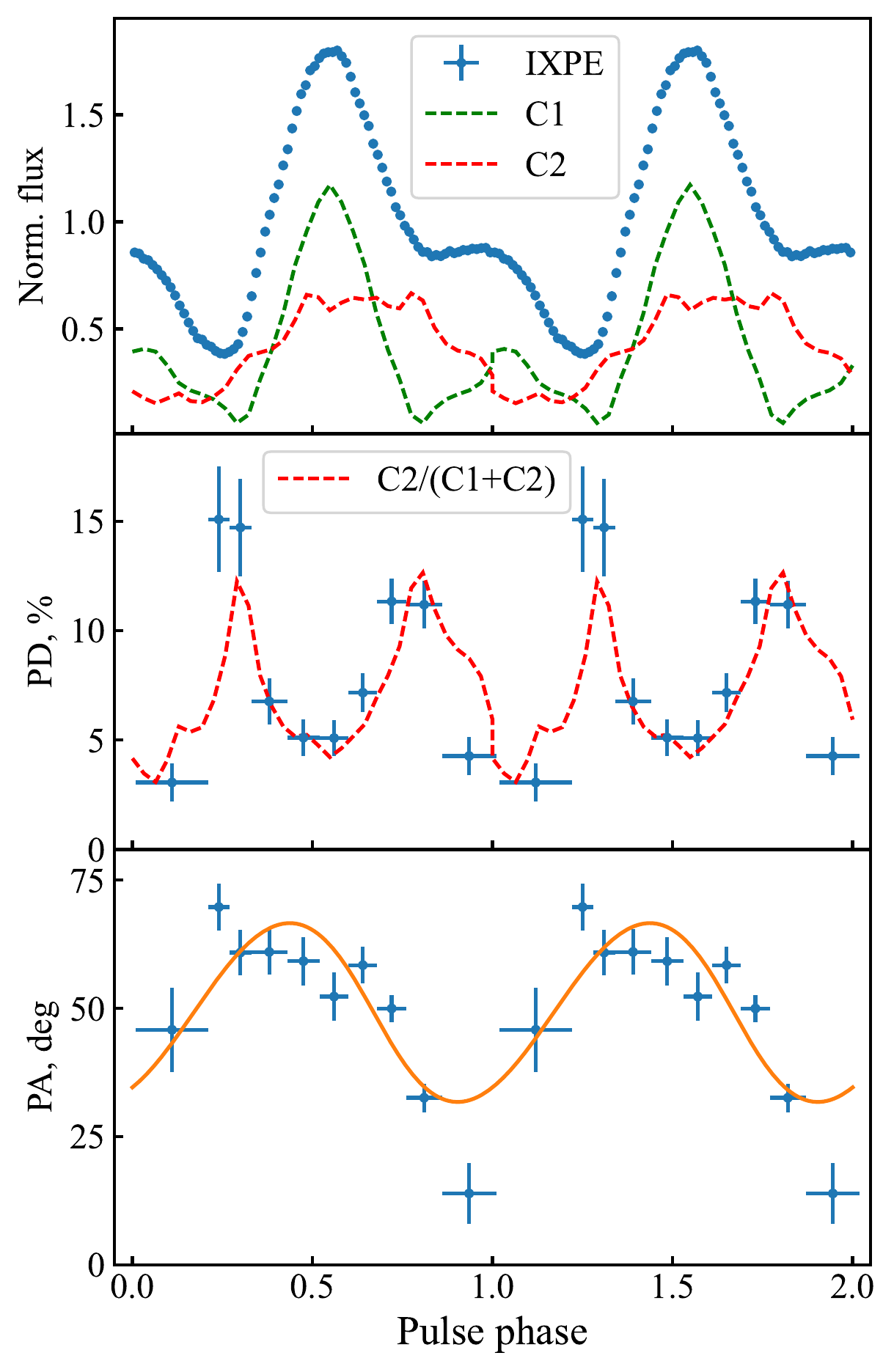}
\caption{{\it Top}: Dependence of the normalized flux in the 2--8 keV energy band (blue points) on the pulse phase in the bright state of \cen. The contribution from two single-pole components C1 and C2 derived by \cite{1996ApJ...467..794K} are shown with green and red dashed lines (see Sect.~\ref{sec:beam_func}). {\it Middle}: Dependence of the PD on phase from the spectro-polarimetric analysis (blue points). Red dashed line represents the relative contributions of one of the poles (C2) to the total flux. {\it Bottom}: Dependence of the PA on the pulse phase. The orange line corresponds to the best-fit rotating vector model (see Sect.~\ref{sec:geom}). }
\label{fig:phres_pd-pa}
\end{figure}
%%%%%%%%%%%%%%%%%%%%%%%%%

%%%%%%%%%%%%%%%%%%%%%%%%%
\begin{table*}
    \caption{Spectral parameters for the phase-resolved spectro-polarimetric analysis in the bright state of \cen}
    \centering
    \begin{tabular}{ccccccc}
    \hline\hline
       Phase  &   $N_{\rm H}$ &  Photon index &  PD & PA   & $\chi^{2}$/d.o.f. \\ 
          &    (10$^{22}$ cm$^{-2}$) &    &  (\%) & (deg)  &   \\\hline
       
0.00--0.20 &  2.9$\pm0.1$  & 1.49$\pm0.02$  & 3.1$\pm0.9$  & 45.8$\pm8.3$  & 1091/1111 \\
0.20--0.26 &  2.5$\pm0.2$  & 1.49$\pm0.06$  & 15.1$\pm2.4$  & 69.8$\pm4.6$  & 997/987 \\
0.26--0.32 &  2.3$\pm0.2$  & 1.35$\pm0.05$  & 14.7$\pm2.2$  & 60.9$\pm4.4$  & 1033/999 \\
0.32--0.42 &  2.9$\pm0.1$  & 1.41$\pm0.02$  & 6.8$\pm1.1$  & 61.0$\pm4.5$  & 1155/1111 \\
0.42--0.51 &  2.9$\pm0.1$  & 1.26$\pm0.02$  & 5.1$\pm0.8$  & 59.2$\pm4.7$  & 1081/1111 \\
0.51--0.59 &  3.0$\pm0.1$  & 1.19$\pm0.02$  & 5.1$\pm0.8$  & 52.3$\pm4.7$  & 1088/1111 \\
0.59--0.67 &  3.4$\pm0.1$  & 1.36$\pm0.02$  & 7.2$\pm0.9$  & 58.4$\pm3.6$  & 1200/1111 \\
0.67--0.75 &  3.4$\pm0.1$  & 1.49$\pm0.02$  & 11.3$\pm1.0$  & 50.0$\pm2.6$  & 1133/1111 \\
0.75--0.85 &  3.1$\pm0.1$  & 1.49$\pm0.03$  & 11.2$\pm1.1$  & 32.6$\pm2.8$  & 1179/1111 \\
0.85--1.00 &  3.2$\pm0.1$  & 1.50$\pm0.02$  & 4.3$\pm0.9$  & 13.9$\pm6.0$  & 1217/1111 \\
    \hline
    \end{tabular}
   \label{tab:fit_phbin}
\end{table*}
%%%%%%%%%%%%%%%%%%%%%%%%%%

In the bright state the spectral shape was found to be slightly different from that in the low one with the iron line flux compatible with zero.
Therefore for the dataset obtained in July 2022 the best-fit model was simplified to {\tt const$\times$tbabs$\times$powerlaw$\times$polconst}.
The quality of the obtained fit can be seen from Figure~\ref{fig:spec-aver}, where the energy spectra for $I$, $Q$ and $U$  are shown along with the residuals. The polarization measurements confidence contours, produced using the {\tt steppar} command in {\sc xspec}, are presented in Figure~\ref{fig:cont-aver}. As it can be seen from the plot, the spectro-polarimetric analysis confirms the discovery of a non-zero polarization signal in the bright state of \cen already found in our previous analysis.

To test the hypothesis of a possible dependence of the polarization properties of \cen on energy, we replaced the {\tt polconst} component of the best-fit model with {\tt pollin} and {\tt polpow}, which correspond to a linear and a power-law dependence of PD and PA on energy, respectively. Application of the modified model to the phase-averaged data resulted in only a marginal improvement of the fit quality ($\Delta \chi^2\sim6.5$ for 2 d.o.f.) with the corresponding F-test probability of 0.065. 

To study the energy spectrum and the polarization as a function of the  spin phase, the data in the bright state have been split into 10 phase bins chosen to guarantee significant measurement of the polarization signal.
Following the approach previously applied to the phase-averaged spectra, the energy spectra for each phase bin has been fitted with the model {\tt const$\times$tbabs$\times$powerlaw$\times$polconst} with the cross-calibration constants fixed at values derived from the phase-averaged analysis, i.e. const$_{\rm DU2}=0.96$ and const$_{\rm DU3}=0.91$. Fit results are summarized in Table~\ref{tab:fit_phbin} and the corresponding confidence contours for PD and PA are presented in Figure~\ref{fig:cont-resol}. 
Variability of the polarization properties of \cen with the pulse phase is presented in Figure~\ref{fig:phres_pd-pa} and is fully compatible with the results obtained from the {\tt pcube} analysis.

In the data collected during the eclipses in July, no significant polarization was measured. 
The corresponding $3\sigma$ upper limit on the PD in the phase-averaged spectrum was obtained with the same best-fit model at the level of 28\%.

\section{Discussion} 
\label{sec:discussion}

XRPs are among the most promising targets for X-ray polarimeters. 
High degree of polarization from these objects was expected due to the strong dependence of the primary processes of radiation and matter interaction such as Compton scattering \citep[e.g.,][]{1986ApJ...309..362D}, free-free and cyclotron absorption and emission \citep[e.g.,][]{2010ApJ...714..630S} on the polarization, energy and direction of X-ray photons.
Birefringence typical for a strongly magnetized plasma allows us to treat the radiative transfer in terms of two normal polarization modes -- the so-called ordinary ``O'' and extraordinary ``X''  \citep{1974JETP...38..903G}. 
Two modes are oriented differently with respect to the plane composed by the magnetic field direction and photons momentum: the electric vector of O-mode photons oscillates within the plane, while the oscillations of electric vector of X-mode photons are  perpendicular to the plane.
Below the cyclotron energy, the opacities of two polarization modes are very different with that of the X-mode being significantly reduced in comparison to the O-mode \citep{2003PhRvL..91g1101L,2006RPPh...69.2631H}.

\begin{figure*}
\centering
\includegraphics[width=0.9\linewidth]{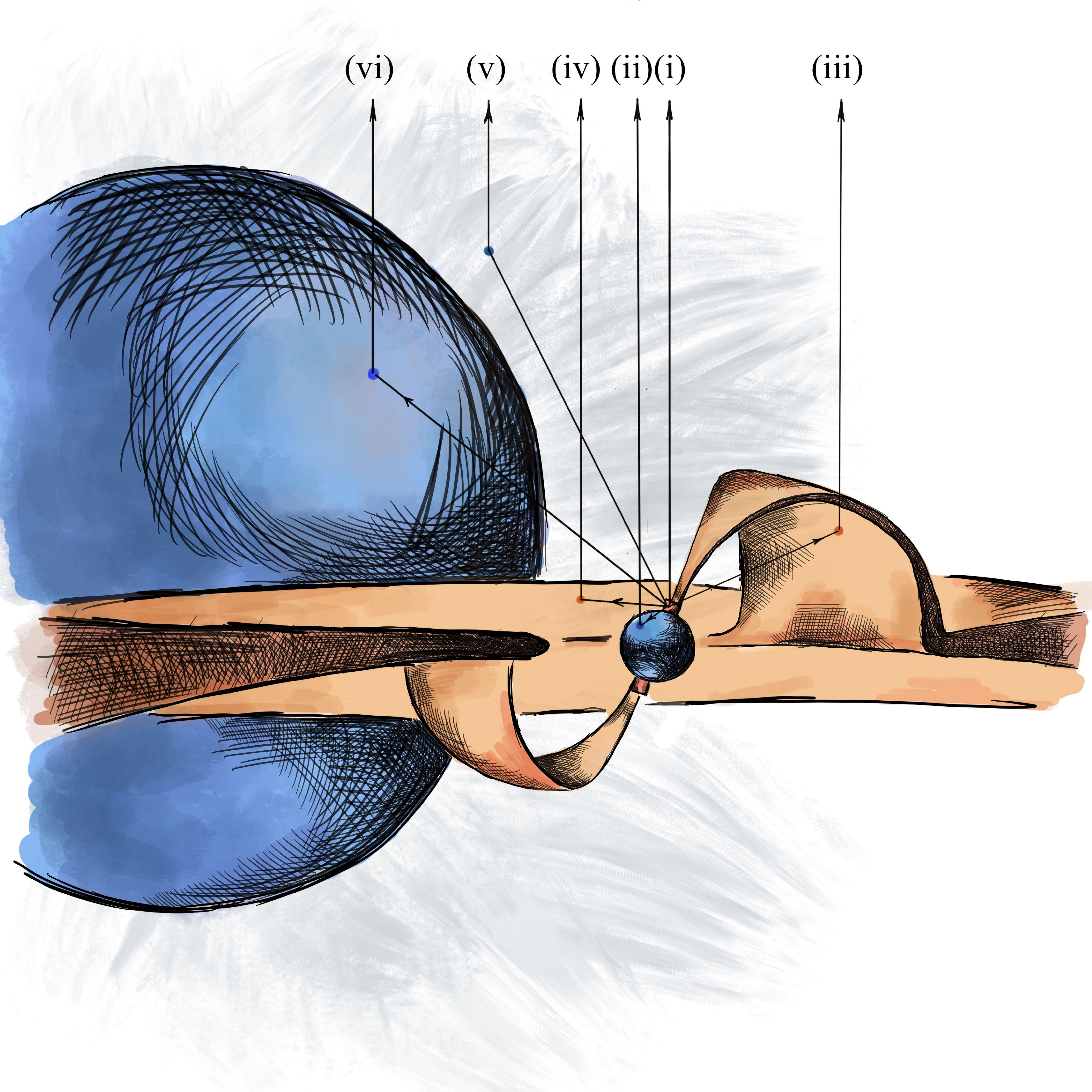}
\caption{Schematic view of the system with the polarization mechanisms considered in this work. Different roman numerals correspond to (i) the intrinsic polarization from the hotspot, (ii) reflection from the NS surface, (iii) reflection from the accretion curtain, (iv) reflection from the accretion disk, (v) scattering by the stellar wind, (vi) reflection by the optical companion.}
 \label{fig:schem}
\end{figure*}

Therefore, the existing models predict the PD from XRPs of up to 80\% (see references in Sect.~\ref{sec:intro}). However, a very complicated interplay of different physical processes and a complex geometrical structure of the accretion flow around XRPs may strongly affect their observed polarization properties. Below we discuss several possible mechanisms (see Figure~\ref{fig:schem}) potentially able to explain a relatively low PD and its evolution with pulse phase in \cen, namely: (i) intrinsic polarization from the hotspot, (ii) reflection of the emission from the NS surface, (iii) reflection from the accretion curtain, (iv)  reflection from the accretion disk, (v) scattering by the stellar wind, and (vi) by the optical companion. In addition, we use our observational results to determine the geometrical parameters of the pulsar.

\subsection{Polarization mechanisms}
\label{sec:mech}

%%%%%%%%%%%%%%%%%%%%%%%%%%
%%%%%%%%%%%%%%%%%%%%%%%%%

\subsubsection{Intrinsic hotspot polarization and the atmospheric properties}

The early theoretical predictions mentioned above are put into question by the relatively low PD of only 5--15\% discovered from \cen in the \ixpe data. However, most of the mentioned models do not take into account temperature structure of the NS atmosphere. Indeed, the observed emission is expected to be dominated by the X- or O-mode depending on structure of the atmosphere, which is determined by the physics of plasma braking in the upper layers of the atmosphere and can be dependent on the mass accretion rate.  The resulting polarization pattern along the pulse profile strongly depends on the configuration of the emission regions, and, correspondingly, on the dominating beam pattern (pencil or fan).

The relatively low PD from \cen is aligned with the recent detection of a  small PD in another XRP, Her~X-1 \citep{Doroshenko2022}.
It appears that the low PD is typical for the sub-critical XRPs, where accretion flow is stopped in the NS atmosphere  \citep{1975AA....42..311B,2012AA...544A.123B,2015MNRAS.447.1847M}.
This result is consistent with a model of accreting NS where the upper atmospheric layers within the hotspots are overheated by the accretion flow \citep{2021MNRAS.503.5193M}.
The critical ingredient of the model is a position of the region where the contributions of plasma and vacuum to the dielectric tensor of magnetized medium become equal, which happens at the mass density $\rho_{\rm V}\sim 10^{-4}\,B_{12}^2 E_{\rm keV}^2\,\,{\rm g\,cm^{-3}}$, where $B_{12}=B/10^{12}$~G is the local magnetic field strength \citep{1979JETP...49..741P,2002ApJ...566..373L}.
The normal modes change their ellipticity passing that region and experience conversion from one to another.
It has been shown by \citet{Doroshenko2022} that the PD of X-ray radiation leaving the atmosphere of a NS can be low in the case when the conversion region is located at the border of the overheated upper layer and the underlying cooler atmosphere.  
PD of the order of 10\% is achieved when the thickness of the heated layer is about 3~g~cm$^{-2}$, which corresponds to the optical depth of about unity, where the contribution to the cooling by free-free emission and Compton scattering is comparable.

\subsubsection{Reflection from the NS surface}
 
A fraction of radiation produced by the hotspots can be scattered within the accretion channel by the free-falling gas. 
The proximity of the \cen luminosity in the bright state to the critical one \citep{1975AA....42..311B,2015MNRAS.447.1847M} should result in a substantial fraction of the hotspot emission to be intercepted by the accreting matter. 
Because of its mildly relativistic velocity, the scattered radiation is beamed towards the NS surface \citep{1988SvAL...14..390L}. 
X-ray radiation intercepted by the atmosphere of a NS is reprocessed and reflected \citep{Poutanen13,2015MNRAS.452.1601P,2021A&A...655A..39K}. 
Reflected radiation is expected to be strongly polarized along the surface with the PD reaching $\sim$45\% in case of a non-magnetic atmosphere \citep{Sobolev63,Gnedin74,Matt93,Poutanen96}. 
For a strongly magnetized NS atmosphere, the differential cross-section is largest for transformation of an O-mode photon to an O-mode one resulting in the dominant polarization parallel to the magnetic field \citep{2022PhRvD.105j3027M}. 
Note, however, that this picture is simplified because the free-free magnetic absorption in the NS atmosphere influences the spectra of reflected radiation in the energy range below $10\,{\rm keV}$.
The X-ray absorption makes the reflected radiation harder \citep{2015MNRAS.452.1601P,2021A&A...655A..39K} and also heats the atmosphere's upper layers.
Heated upper layers can contribute to the X-ray spectra and polarization in the considered energy range. Thus, X-ray spectra and polarization analyzes require the self-consistent calculation of X-ray reflection from the atmosphere and account for the atmospheric temperature structure influenced by external illumination.
We suggest, however, that reflection from the surface can contribute significantly to the observed polarimetric signal.

\subsubsection{Reflection from the accretion curtain}

The potential effect of the matter flow from the accretion disk towards the NS surface, the so-called accretion curtain, on the pulse profile formation and polarization properties was recognised long time ago \citep{1976SvAL....2..111S,BaskoSunyaev76SvA}.
Similarly to the mechanism discussed above for the NS surface, the emission reflected from a curtain is expected to be highly polarized. 
Moreover, the magnetic field at large distance from the NS does not play a significant role and non-magnetic Thomson scattering operates.
In the bright state of \cen ($L\sim 3\times 10^{37}\,{\rm erg\,s^{-1}}$), the co-latitude of the polar cap edge is expected to be $\theta_{\rm pol}\sim 4\degr$.
The optical thickness of accretion flow due to Thomson scattering across the field lines is $\sim 5$ at the NS surface and decreases rapidly with the distance and corresponding co-latitude.
Calculating the dynamics of accretion flow along dipole magnetic field lines under the influence of gravitational and centrifugal forces \citep{2017MNRAS.467.1202M} we get approximate dependence of the flow optical thickness on the co-latitude in the bright state of \cen: $\tau_{\rm T}\approx 5\,(\theta_{\rm pol}/\theta)^2$.
This approximation is valid for $\theta\lesssim20\degr$.
For $\theta\lesssim8\degr$, the Thomson optical thickness of the flow is above unity and, thus, the X-ray radiation leaving hot spots within this small angle to the magnetic dipole can be significantly modified by scattering in the accretion channel.
Assuming the intensity of X-ray radiation to be independent on the direction at the NS surface, we get that $\sim 2\%$ of the total luminosity can be intercepted and reprocessed by the accretion curtain above the NS surface. 
As a result, we expect that  scattering of the X-ray radiation by material covering the Alfv\'en surface can produce polarization within a couple of per cents, which is not enough to explain the whole observed signal.

\subsubsection{Reflection from the accretion  disk}

In addition to the accretion curtain, reflection may occur from the accretion disk. 
For a mass ratio $q=M_{\rm NS}/M_{\rm O}\approx0.059$ and the orbit size  $a=19R_{\odot}$, the size of the Roche lobe around the NS is about \citep{Paczynski71} $R_{\rm RL}\approx 0.462 q^{1/3} a \approx  2.4\times10^{11}$~cm. 
The accretion disk size, limited by the tidal forces, is then about half of that \citep{Blondin00}, i.e. $\sim$4 lt-s. 
This is large enough to nearly smear out variations of the flux reflected from the (flared) disk with the 4.8-s pulsar spin phase.  
The fraction of X-ray photons intercepted by the disk depends mostly on its opening angle and is estimated to be about 0.2 \citep{Verbunt99}. 
The X-ray photons impinging on the accretion disk at a nearly tangential direction and reflected to the observer at inclination of $\sim$70\degr\ are polarized at a level of $\sim$30\% nearly independent of energy \citep[see Figure~5 in][]{Poutanen96}. 
Accounting for the fraction of intercepted photons of 0.2 and a strong dependence of the reflection albedo on energy (which is about  10\% at 5 keV), we get PD$\lesssim1\%$ in the IXPE range (rather constant with the spin phase).  
Thus, reflection from the disk cannot contribute significantly to the observed polarization.

\subsubsection{Scattering by the wind}

As was demonstrated in several studies, the NS in \cen is embedded in a dense inhomogeneous stellar wind \citep{Sanjurjo-Ferrin21}.
Let us roughly estimate the influence of stellar wind on pulsar's X-ray polarization.
If the mass-loss rate due to the stellar wind is $\dot{M}_{\rm w}$, the wind is spherically symmetric and has a constant velocity $v_{\rm w}$, the mass density of the wind at a separation $a$ is $\rho_{\rm w}=\dot{M}_{\rm w}/(4\pi a^2 v_{\rm w})$.
Then the column mass density of the wind material can be estimated as $\Sigma\sim \int_{a}^{\infty}{\rm d}r\,\rho_0 (a/r)^2=\rho_0 a$.
Taking $a\approx 19R_{\odot}$, $\dot{M}_{\rm w}\approx 5\times 10^{-7}\,M_\odot\,{\rm yr^{-1}}$ \citep{2019A&A...621A..85H,Sanjurjo-Ferrin21},
and $v_w\sim 1000$\,km\,s$^{-1}$, we get the wind mass density at the NS orbit $\rho_0\approx 1.4\times 10^{-14}\,{\rm g\,cm^{-3}}$ and the typical column mass density $\Sigma\sim 0.02\,{\rm g\,cm^{-2}}$ corresponding to $\tau_{\rm T}<0.01$.
Thus, because of the small optical thickness for scattering and a more or less spherical geometry of the wind around the NS, the stellar wind is not a likely source of the observed polarization.

\subsubsection{Scattering from the optical companion}

The optical companion, on the other hand, occupies a rather large part of the sky, $\Omega/(4\pi)=\frac{1}{4} (R_{\rm O}/a)^2\approx0.1$, as seen from the pulsar. 
Thus, a substantial fraction of the X-ray light emitted by the pulsar can be reflected from the donor star, depending on the emission pattern of the pulsar. 
This radiation is polarized \citep{Basko74,Gnedin74}. 
However, most of this radiation escapes at energies above 10 keV, because at lower energies photons will more likely be absorbed photo-electrically. 
At high energies, in a pure scattering case, polarization of the scattered light reaches about 33\%  \citep[e.g,][]{Buenzli09} when the donor star is in quadratures, producing about PD$\approx$3\%  of the total light. 
At lower energies, absorption in the stellar atmosphere will dominate and reflected photons are scattered only once. This increases the PD of the scattered component, but reduced the fraction of scattered photons. 
Thus, the reflection from the donor star can produce a weak polarization signal at the highest end of the \ixpe range at the level of PD$\sim$1\% varying with the orbital phase. 
Because of a $\sim$16--44~lt-s separation between the stellar surface and the 4.8-s pulsar, reflection from the star obviously cannot produce spin-dependent polarization. 

\subsection{Geometry of the system}
\label{sec:geom}

%%%%%%%%%%%%%%%%%%%%%%%%%%%%
\begin{figure*}
\centering
\includegraphics[width=0.85\linewidth]{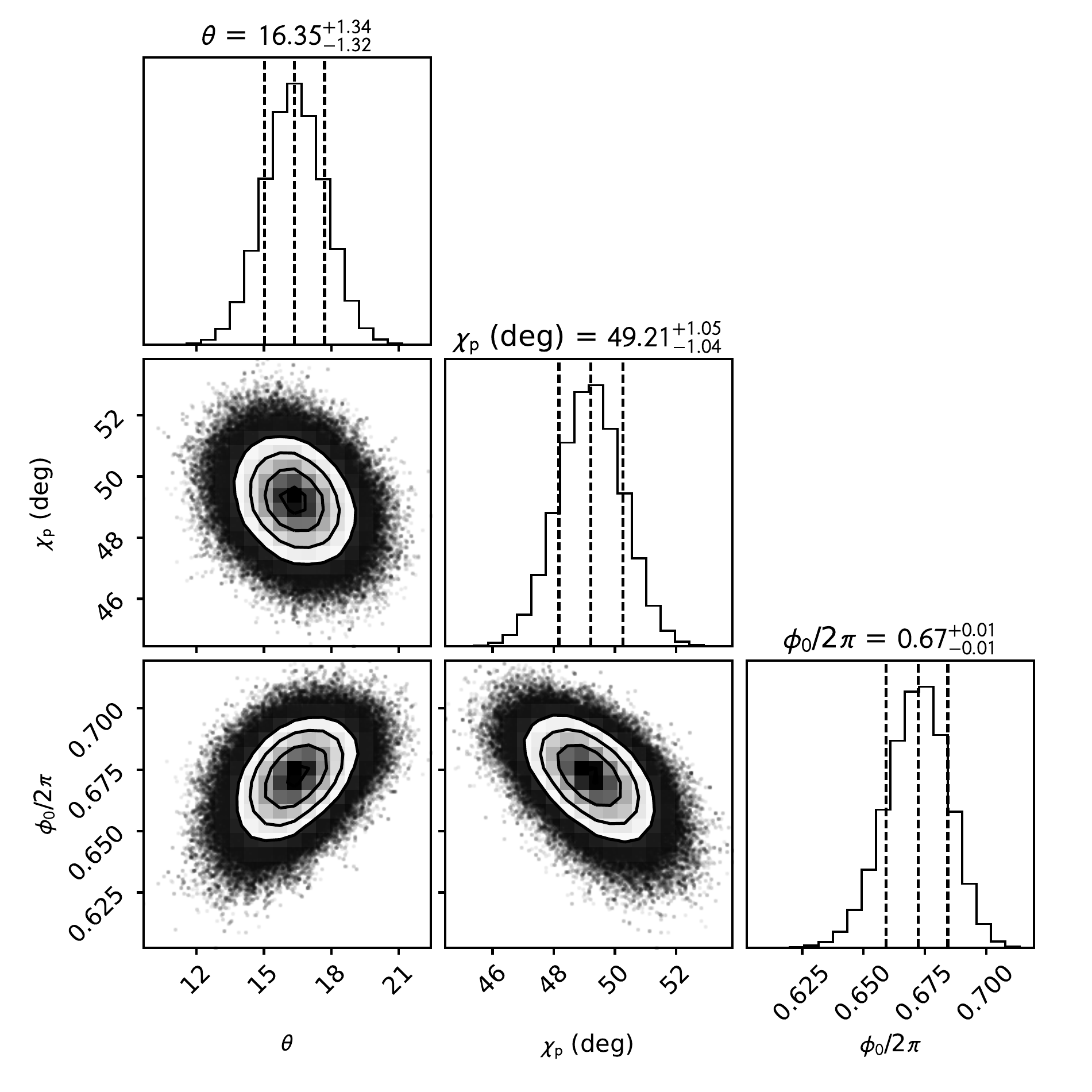}
\caption{Corner plot of posterior distribution for the RVM parameters for the pulsar inclination angle $i_{\rm p}$ fixed at the orbital inclination value of $70\fdg2$. The contours correspond to two-dimensional 1, 2, and 3$\sigma$ confidence levels. }
\label{fig:emcee_i70}
\end{figure*}
%%%%%%%%%%%%%%%%%%%%%%%%%%%%

Because \cen is an X-ray pulsar it is natural to expect the orientation of the polarization to change with the rotational phase.
Although the magnetic field near the star and near the accretion disk is probably complicated, in between it is expected to be approximately dipolar.  
In fact this assumption is the basis to understand the physics of the interaction of the magnetic field and the accretion disk that channels material onto the polar region of the surface of the NS, resulting in the pulsed emission \citep[e.g.][]{1979ApJ...234..296G}. 
\citet{1978PAZh....4..214G} argued that vacuum birefringence causes the  radiation to propagate in the magnetosphere in two normal modes. This propagation in the normal modes can continue until the polarization-limiting radius \citep{1952RSPSA.215..215B,2002PhRvD..66b3002H,2018Galax...6...76H} which is about twenty stellar radii (300~km), much larger than the star and also much smaller than the inner edge of the accretion disk; that is where we expect the field configuration to be dipolar.
Thus the final polarization of radiation measured at the telescope is parallel or perpendicular to projection of  the instantaneous magnetic axis of the star onto the plane of the sky.  
Under these assumptions, the rotating-vector model (RVM) of \citet{1969ApL.....3..225R} holds to a good approximation \citep[see also][]{unbinned}.  
Because the observed direction of the polarization is generated far from the stellar surface, the evolution of the polarization in phase does not have to coincide with the evolution of the flux.

In the RVM,  if radiation escapes in the O-mode the PA can be described by the following expression  \citep{Poutanen20RVM,Doroshenko2022} 
\begin{equation} \label{eq:pa_rvm}
\tan (\mbox{PA}\!-\!\chi_{\rm p})\!=\! \frac{-\sin \theta\ \sin (\phi-\phi_0)}
{\sin i_{\rm p} \cos \theta\!  - \! \cos i_{\rm p} \sin \theta  \cos (\phi\!-\!\phi_0) } .
\end{equation} 
Here $\chi_{\rm p}$ is the position angle of the pulsar spin, $i_{\rm p}$ is the pulsar inclination (i.e. the angle between the pulsar spin vector and the line-of-sight), $\theta$ is the magnetic obliquity (i.e. the angle between the magnetic dipole and the spin axes), $\phi_0$ is the phase when the emission region is closest to the observer, and $\phi$ is the pulse phase.
  
We fit the RVM to the pulse phase dependence of the PA obtained from the spectro-polarimetric analysis using the  affine invariant Markov chain Monte Carlo ensemble sampler {\sc emcee} package of {\sc python} \citep{2013PASP..125..306F}. The available data do not allow to constrain $i_{\rm p}$ and, therefore, we fixed it to the orbital inclination $i_{\rm p}=i_{\rm orb}=70\fdg2$ \citep{1999MNRAS.307..357A}. This resulted in accurate estimates of the co-latitude of the magnetic pole, $\theta=16\fdg4\pm1\fdg3$, and the position angle of the pulsar spin, $\chi_{\rm p}=\chi_{\rm p,O}=49\fdg2\pm1\fdg1$ (see Figures~\ref{fig:phres_pd-pa} and \ref{fig:emcee_i70}). It is important to mention that because only the orientation of the polarization plane can be measured, the pulsar spin can be oppositely directed at $\chi_{\rm p}=\chi_{\rm p,O}+180\degr=209\fdg2\pm1\fdg1$. If radiation escapes in the X-mode, then the pulsar spin is oriented at $\chi_{\rm p}=\chi_{\rm p,X}=\chi_{\rm p,O}\pm90\degr$.
The uncertainties in the direction of the spin and the intrinsic polarization mode have no effect on the best-fit $\theta$. 

Because it is not known whether the orbit is clock- or anticlockwise, we have also  considered an inclination of $i_{\rm p}=180\degr-i_{\rm orb}=109\fdg8$.  
The best-fit parameters  $\theta=18\fdg6\pm1\fdg4$ and $\chi_{\rm p}=47\fdg7\pm1\fdg0$
do not differ much from those  obtained for $i_{\rm p}=i_{\rm orb}$, because the contribution of the cosine term in the denominator of Eq.~(\ref{eq:pa_rvm}) is small for close to edge-on inclinations and small amplitude of PA (implying small $\theta$). 
Therefore, the impact of changing $i_{\rm p}=i_{\rm orb}$ to $i_{\rm p}=180\degr-i_{\rm orb}$ is small. 

The results were verified using the unbinned polarimetric analysis when the RVM is fitted to the measured Stokes parameters corrected for 
spurious modulation on a photon-by-photon basis as outlined in \citet{unbinned} and \citet{2021AJ....162..134M}. 
The obtained PD and PA are nearly identical to those shown in Figure~\ref{fig:phres_pd-pa}.

\subsection{Beam function}
\label{sec:beam_func}

The determined geometrical parameters of the pulsar and the phase behavior of its polarimetric properties agree surprisingly well with the pulse profile decomposition into two single-pole components performed by \cite{1996ApJ...467..794K}. These authors demonstrated that the pulse profile of \cen is compatible with a slightly displaced from the antipodal positions (by approximately 10\degr) dipole geometry with the co-latitude of the main component of $\theta=18\degr$. As it can be seen from  Figure~\ref{fig:emcee_i70}, this value is well compatible with our polarimetric analysis. Another fact pointing to the possible correctness of the profile decomposition presented by \cite{1996ApJ...467..794K} is the correlation of the PD with the relative contribution of one of the poles (C2) to the total flux (see Figure~\ref{fig:phres_pd-pa}). Indeed, one can see that the PD reaches a minimum at phases where the main peak (C1) is dominating. This can be understood if this component appears due to pencil beam emission diagram. Indeed, it was shown by \cite{1988ApJ...324.1056M}, that in the case of sub-critical accretion \citep{2015MNRAS.447.1847M} when pencil beam diagram naturally appears, one would expect an anti-correlation between the pulsed flux and PD. In this case the second component of the profile (C2) may correspond to the antipodal hotspot seen at a large angle. This picture may at least partly explain the relatively low PD by mixing of emission from two poles seen at different angles. Indeed, \cite{Doroshenko2022} showed that emission emerging from a heated atmosphere can be dominated by X- or O-mode depending on the zenith angle. Being polarized perpendicular to each other a complex interplay of the two modes throughout the rotation cycle would lead to a significant  decrease of the PD. Physically this fan-beamed component may originate from the emission reflected either from the NS surface or an accretion curtain as discussed above.

\section{Summary}
\label{sec:sum}

The results of our study can be summarized as follows:

\begin{enumerate}

\item \cen was observed by \ixpe twice over the periods of 2022
Jan 29--31 and July 4--7 in the low and bright states respectively, when the off-eclipse source flux was different by a factor of $\sim20$.

\item Both the energy-binned and the spectro-polarimetric  analyses of the phase-averaged data in the 2--8 keV band revealed a significant polarization of the source in the bright state with the PD of 5.8$\pm$0.3\% and the PA of $49\fdg6\pm1\fdg5$. In the low-luminosity state, no significant polarization was found.

\item In the phase-resolved data collected in the bright state, a significant anti-correlation between the flux and the PD, as well as a strong variation of the PA, were discovered. One of the single-pole components from the pulse profile decomposition by \citet{1996ApJ...467..794K} was found to dominate the polarization signal over the pulse phase. 

\item We obtained a solution for the geometrical parameters of the pulsar applying a rotating-vector model. The fit resulted in a position angle of the pulsar angular momentum of about 49\degr\ (or 209\degr) if radiation escapes from the surface in the O-mode, or 139\degr\ (or $-$41\degr) in case of the X-mode emission. 
In all cases, the magnetic obliquity was found to be rather low $\sim$16\degr. 

\item The relatively low polarization detected from \cen can be explained in the framework of the NS atmosphere model with the upper layers overheated by the accreted matter. Another possible reason is mixing of emission from two magnetic poles seen at different angles. 

\item A fraction of the detected polarization signal may come from the reflection of radiation scattered in the accretion channel from the NS surface. 
Also, reflection by the accretion curtain may contribute to the observed polarization.

\end{enumerate}

\begin{acknowledgments}
The Imaging X-ray Polarimetry Explorer (IXPE) is a joint US and Italian mission.  The US contribution is supported by the National Aeronautics and Space Administration (NASA) and led and managed by its Marshall Space Flight Center (MSFC), with industry partner Ball Aerospace (contract NNM15AA18C).  
The Italian contribution is supported by the Italian Space Agency (Agenzia Spaziale Italiana, ASI) through contract ASI-OHBI-2017-12-I.0, agreements ASI-INAF-2017-12-H0 and ASI-INFN-2017.13-H0, and its Space Science Data Center (SSDC), and by the Istituto Nazionale di Astrofisica (INAF) and the Istituto Nazionale di Fisica Nucleare (INFN) in Italy.  This research used data products provided by the IXPE Team (MSFC, SSDC, INAF, and INFN) and distributed with additional software tools by the High-Energy Astrophysics Science Archive Research Center (HEASARC), at NASA Goddard Space Flight Center (GSFC).

We acknowledge support from the Russian Science Foundation grant \mbox{20-12-00364} (SST, JP, VFS), the Academy of Finland grants 333112, 349144, 349373, and 349906 (SST, JP), the German Academic Exchange Service (DAAD) travel grant 57525212 (VD, VFS), the German Research Foundation (DFG) grant \mbox{WE 1312/53-1} (VFS), UKRI Stephen Hawking fellowship and the Netherlands Organization for Scientific Research Veni fellowship (AAM). 
\end{acknowledgments}


\vspace{5mm}
\facilities{\ixpe}


\software{{\sc astropy} \citep{2013A&A...558A..33A,2018AJ....156..123A}, 
{\sc xspec} \citep{Arn96}, 
{\sc ixpeobssim} \citep{2022arXiv220306384B}, 
 {\sc emcee} \citep{2013PASP..125..306F}
          }



%\clearpage 

\appendix

\section{Timing analysis and orbital parameters}
\label{app:1}

A coherent pulsar timing solution is required in order to conduct phase-resolved polarimetric analysis, and the fact that IXPE observation cover significant fraction of the orbital cycle (particularly the second observation) implies that correction for motion of the pulsar in the binary system is essential to obtain such a solution. Cen~X-3 is one of the best studied accreting pulsars and orbital parameters of the system are known with most recent estimate published by \cite{2010MNRAS.401.1532R}. We found, however, that extrapolating these ephemerides based on the RXTE observations of the source in 1997 to current date results in residual regular variations of the observed spin frequency and pulse arrival times even after correction, which is mostly related to accumulated error in estimated mid-eclipse ($T_{90}$) time. We emphasize that our goal here is not to obtain updated orbital ephemerides but merely phase IXPE data, and more sophisticated analysis of the orbit will be published elsewhere. Nevertheless, in order to improve the orbital solution we conducted, therefore, pulsar timing analysis following an approach largely similar to that by \citet{2010MNRAS.401.1532R} as described below. 

\begin{figure}
\centering
\includegraphics[width=0.7\columnwidth]{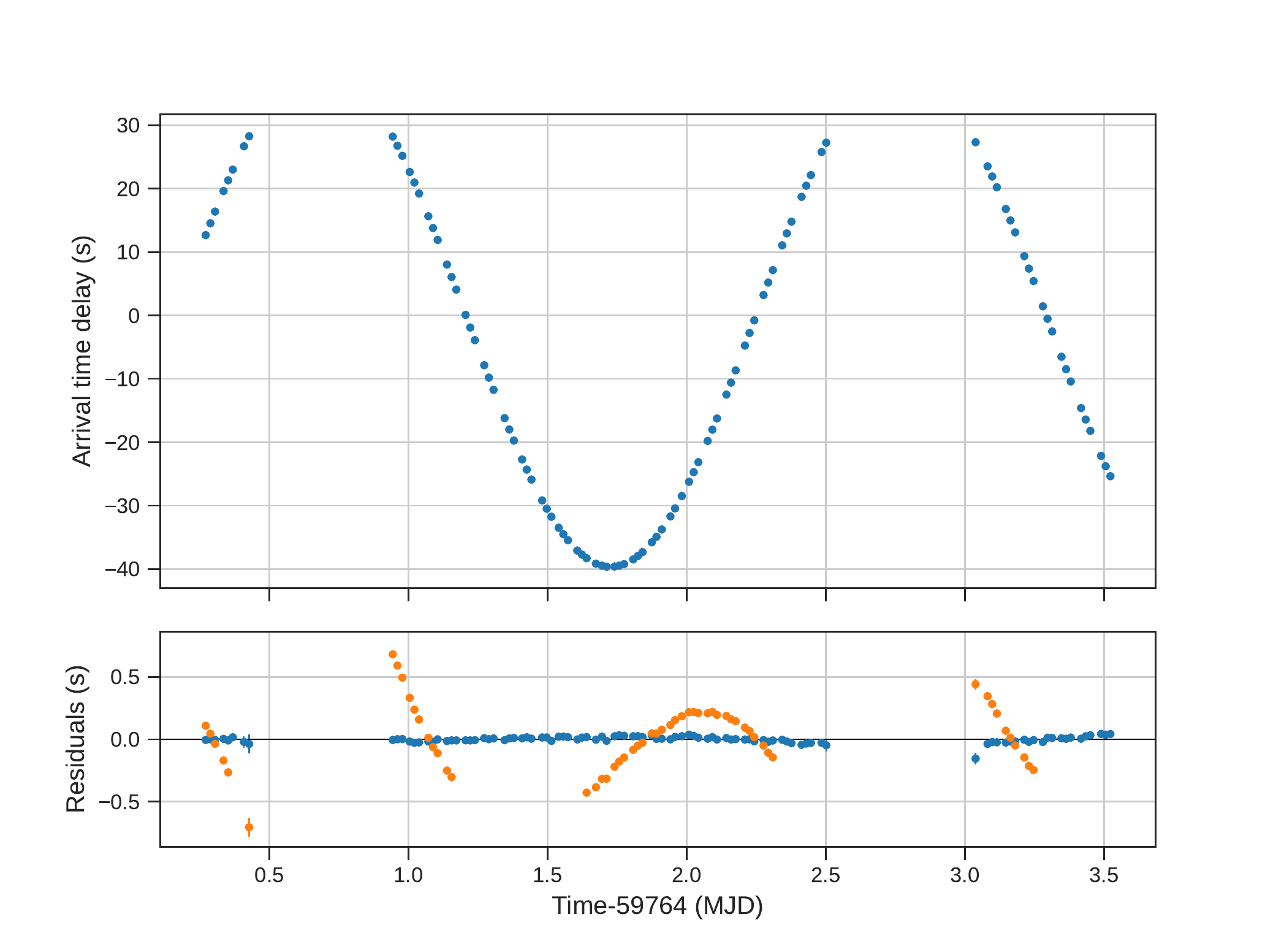}
\caption{Top: pulse time arrival delays with respect to constant period. 
Bottom: residuals for the best-fit models  assuming ephemerides by \cite{2010MNRAS.401.1532R} in orange and the final timing solution obtained in this work for the second IXPE observation in blue.}
\label{fig:toa_resid}
\end{figure}

After correcting the observed event times to the Solar system barycenter, we divided the observation in set of intervals corresponding to good time intervals of the observation (mostly defined by SAA passages), and determined the pulsation frequency in each of those intervals. We then roughly estimated the $T_{90}$ value based on the resulting sinusoidal modulation of the observed spin frequency, which was used as a starting point for the subsequent analysis. In particular, we corrected the light-curve using ephemerides by \citet{2010MNRAS.401.1532R} with adjusted $T_{90}$. After that we searched for pulsations, and folded the corrected light curve with the source spin period in order to obtain a high-quality pulse profile template. This template was then used to determine pulse times of arrival (TOAs) in the uncorrected light curve by direct fitting of the template pulse profile to the local pulse profiles estimated for a set of time intervals containing at least 150 pulses ($\sim$12\,min, where each interval was folded using the frequency estimated based on the initial ephemerides described above). 

The obtained pulse arrival times were then modeled assuming a circular orbit in a same way as done by \citet{2010MNRAS.401.1532R} but only considering $T_{90}$, spin frequency and spin frequency derivative as free parameters. As already discussed by \citet{2010MNRAS.401.1532R}, timing noise likely associated with pulse profile variations implies that it is not possible to obtain fully regular residuals with this approach. We incorporated an additional systematic error of 0.0178\,s in order to achieve statistically acceptable fit with reduced $\chi^2_{\rm red}\sim1$, which is necessary to estimate uncertainties for model parameters. As a result we find $T_{90}=59764.68380(4)$~MJD, $P_{\rm spin}=4.7957473(8)$\,s and $\dot{P}_{\rm spin}=-1.45(5)\times10^{-10}$\,s\,s$^{-1}$ (epoch corresponds to the first pulse arrival time, i.e. MJD~59764.27039373). The best-fit residuals both for ephemerides by \citet{2010MNRAS.401.1532R} and those obtained in this work are presented in Figure~\ref{fig:toa_resid}.


\bibliography{allbib}{}

\begin{thebibliography}{}
\expandafter\ifx\csname natexlab\endcsname\relax\def\natexlab#1{#1}\fi
\providecommand{\url}[1]{\href{#1}{#1}}
\providecommand{\dodoi}[1]{doi:~\href{http://doi.org/#1}{\nolinkurl{#1}}}
\providecommand{\doeprint}[1]{\href{http://ascl.net/#1}{\nolinkurl{http://ascl.net/#1}}}
\providecommand{\doarXiv}[1]{\href{https://arxiv.org/abs/#1}{\nolinkurl{https://arxiv.org/abs/#1}}}

\bibitem[{{Abarr} {et~al.}(2020){Abarr}, {Baring}, {Beheshtipour}, {Beilicke},
  {de Geronimo}, {Dowkontt}, {Errando}, {Guarino}, {Iyer}, {Kislat}, {Kiss},
  {Kitaguchi}, {Krawczynski}, {Lanzi}, {Li}, {Lisalda}, {Okajima}, {Pearce},
  {Press}, {Rauch}, {Stuchlik}, {Takahashi}, {Tang}, {Uchida}, {West}, {Jenke},
  {Krimm}, {Lien}, {Malacaria}, {Miller}, \&
  {Wilson-Hodge}}]{2020ApJ...891...70A}
{Abarr}, Q., {Baring}, M., {Beheshtipour}, B., {et~al.} 2020, \apj, 891, 70,
  \dodoi{10.3847/1538-4357/ab672c}

\bibitem[{{Arnason} {et~al.}(2021){Arnason}, {Papei}, {Barmby}, {Bahramian}, \&
  {Gorski}}]{2021MNRAS.502.5455A}
{Arnason}, R.~M., {Papei}, H., {Barmby}, P., {Bahramian}, A., \& {Gorski},
  M.~D. 2021, \mnras, 502, 5455, \dodoi{10.1093/mnras/stab345}

\bibitem[{{Arnaud}(1996)}]{Arn96}
{Arnaud}, K.~A. 1996, in ASP Conf. Ser., Vol. 101, Astronomical Data Analysis
  Software and Systems V, ed. G.~H. {Jacoby} \& J.~{Barnes} (San Francisco:
  Astron. Soc. Pac.), 17--20

\bibitem[{{Ash} {et~al.}(1999){Ash}, {Reynolds}, {Roche}, {Norton}, {Still}, \&
  {Morales-Rueda}}]{1999MNRAS.307..357A}
{Ash}, T.~D.~C., {Reynolds}, A.~P., {Roche}, P., {et~al.} 1999, \mnras, 307,
  357, \dodoi{10.1046/j.1365-8711.1999.02605.x}

\bibitem[{{Astropy Collaboration} {et~al.}(2013){Astropy Collaboration},
  {Robitaille}, {Tollerud}, {Greenfield}, {Droettboom}, {Bray}, {Aldcroft},
  {Davis}, {Ginsburg}, {Price-Whelan}, {Kerzendorf}, {Conley}, {Crighton},
  {Barbary}, {Muna}, {Ferguson}, {Grollier}, {Parikh}, {Nair}, {Unther},
  {Deil}, {Woillez}, {Conseil}, {Kramer}, {Turner}, {Singer}, {Fox}, {Weaver},
  {Zabalza}, {Edwards}, {Azalee Bostroem}, {Burke}, {Casey}, {Crawford},
  {Dencheva}, {Ely}, {Jenness}, {Labrie}, {Lim}, {Pierfederici}, {Pontzen},
  {Ptak}, {Refsdal}, {Servillat}, \& {Streicher}}]{2013A&A...558A..33A}
{Astropy Collaboration}, {Robitaille}, T.~P., {Tollerud}, E.~J., {et~al.} 2013,
  \aap, 558, A33, \dodoi{10.1051/0004-6361/201322068}

\bibitem[{{Astropy Collaboration} {et~al.}(2018){Astropy Collaboration},
  {Price-Whelan}, {Sip{\H{o}}cz}, {G{\"u}nther}, {Lim}, {Crawford}, {Conseil},
  {Shupe}, {Craig}, {Dencheva}, {Ginsburg}, {VanderPlas}, {Bradley},
  {P{\'e}rez-Su{\'a}rez}, {de Val-Borro}, {Aldcroft}, {Cruz}, {Robitaille},
  {Tollerud}, {Ardelean}, {Babej}, {Bach}, {Bachetti}, {Bakanov}, {Bamford},
  {Barentsen}, {Barmby}, {Baumbach}, {Berry}, {Biscani}, {Boquien}, {Bostroem},
  {Bouma}, {Brammer}, {Bray}, {Breytenbach}, {Buddelmeijer}, {Burke},
  {Calderone}, {Cano Rodr{\'\i}guez}, {Cara}, {Cardoso}, {Cheedella}, {Copin},
  {Corrales}, {Crichton}, {D'Avella}, {Deil}, {Depagne}, {Dietrich}, {Donath},
  {Droettboom}, {Earl}, {Erben}, {Fabbro}, {Ferreira}, {Finethy}, {Fox},
  {Garrison}, {Gibbons}, {Goldstein}, {Gommers}, {Greco}, {Greenfield},
  {Groener}, {Grollier}, {Hagen}, {Hirst}, {Homeier}, {Horton}, {Hosseinzadeh},
  {Hu}, {Hunkeler}, {Ivezi{\'c}}, {Jain}, {Jenness}, {Kanarek}, {Kendrew},
  {Kern}, {Kerzendorf}, {Khvalko}, {King}, {Kirkby}, {Kulkarni}, {Kumar},
  {Lee}, {Lenz}, {Littlefair}, {Ma}, {Macleod}, {Mastropietro}, {McCully},
  {Montagnac}, {Morris}, {Mueller}, {Mumford}, {Muna}, {Murphy}, {Nelson},
  {Nguyen}, {Ninan}, {N{\"o}the}, {Ogaz}, {Oh}, {Parejko}, {Parley}, {Pascual},
  {Patil}, {Patil}, {Plunkett}, {Prochaska}, {Rastogi}, {Reddy Janga},
  {Sabater}, {Sakurikar}, {Seifert}, {Sherbert}, {Sherwood-Taylor}, {Shih},
  {Sick}, {Silbiger}, {Singanamalla}, {Singer}, {Sladen}, {Sooley},
  {Sornarajah}, {Streicher}, {Teuben}, {Thomas}, {Tremblay}, {Turner},
  {Terr{\'o}n}, {van Kerkwijk}, {de la Vega}, {Watkins}, {Weaver}, {Whitmore},
  {Woillez}, {Zabalza}, \& {Astropy Contributors}}]{2018AJ....156..123A}
{Astropy Collaboration}, {Price-Whelan}, A.~M., {Sip{\H{o}}cz}, B.~M., {et~al.}
  2018, \aj, 156, 123, \dodoi{10.3847/1538-3881/aabc4f}

\bibitem[{{Baldini} {et~al.}(2022){Baldini}, {Bucciantini}, {Di Lalla},
  {Ehlert}, {Manfreda}, {Omodei}, {Pesce-Rollins}, \&
  {Sgr{\`o}}}]{2022arXiv220306384B}
{Baldini}, L., {Bucciantini}, N., {Di Lalla}, N., {et~al.} 2022, arXiv
  e-prints, arXiv:2203.06384.
\newblock \doarXiv{2203.06384}

\bibitem[{{Baldini} {et~al.}(2021){Baldini}, {Barbanera}, {Bellazzini},
  {Bonino}, {Borotto}, {Brez}, {Caporale}, {Cardelli}, {Castellano},
  {Ceccanti}, {Citraro}, {Di Lalla}, {Latronico}, {Lucchesi}, {Magazz{\`u}},
  {Magazz{\`u}}, {Maldera}, {Manfreda}, {Marengo}, {Marrocchesi}, {Mereu},
  {Minuti}, {Mosti}, {Nasimi}, {Nuti}, {Oppedisano}, {Orsini}, {Pesce-Rollins},
  {Pinchera}, {Profeti}, {Sgr{\`o}}, {Spandre}, {Tardiola}, {Zanetti}, {Amici},
  {Andersson}, {Attin{\`a}}, {Bachetti}, {Baumgartner}, {Brienza},
  {Carpentiero}, {Castronuovo}, {Cavalli}, {Cavazzuti}, {Centrone}, {Costa},
  {D'Alba}, {D'Amico}, {Del Monte}, {Di Cosimo}, {Di Marco}, {Di Persio},
  {Donnarumma}, {Evangelista}, {Fabiani}, {Ferrazzoli}, {Kitaguchi}, {La
  Monaca}, {Lefevre}, {Loffredo}, {Lorenzi}, {Mangraviti}, {Matt}, {Meilahti},
  {Morbidini}, {Muleri}, {Nakano}, {Negri}, {Nenonen}, {O'Dell}, {Perri},
  {Piazzolla}, {Pieraccini}, {Pilia}, {Puccetti}, {Ramsey}, {Rankin},
  {Ratheesh}, {Rubini}, {Santoli}, {Sarra}, {Scalise}, {Sciortino}, {Soffitta},
  {Tamagawa}, {Tennant}, {Tobia}, {Trois}, {Uchiyama}, {Vimercati},
  {Weisskopf}, {Xie}, {Zanetti}, \& {Zhou}}]{2021APh...13302628B}
{Baldini}, L., {Barbanera}, M., {Bellazzini}, R., {et~al.} 2021, Astroparticle
  Physics, 133, 102628, \dodoi{10.1016/j.astropartphys.2021.102628}

\bibitem[{{Basko} \& {Sunyaev}(1975)}]{1975AA....42..311B}
{Basko}, M.~M., \& {Sunyaev}, R.~A. 1975, \aap, 42, 311

\bibitem[{{Basko} \& {Sunyaev}(1976)}]{BaskoSunyaev76SvA}
---. 1976, \sovast, 20, 537

\bibitem[{{Basko} {et~al.}(1974){Basko}, {Sunyaev}, \& {Titarchuk}}]{Basko74}
{Basko}, M.~M., {Sunyaev}, R.~A., \& {Titarchuk}, L.~G. 1974, \aap, 31, 249

\bibitem[{{Becker} {et~al.}(2012){Becker}, {Klochkov}, {Sch{\"o}nherr},
  {Nishimura}, {Ferrigno}, {Caballero}, {Kretschmar}, {Wolff}, {Wilms}, \&
  {Staubert}}]{2012AA...544A.123B}
{Becker}, P.~A., {Klochkov}, D., {Sch{\"o}nherr}, G., {et~al.} 2012, \aap, 544,
  A123, \dodoi{10.1051/0004-6361/201219065}

\bibitem[{{Blondin}(2000)}]{Blondin00}
{Blondin}, J.~M. 2000, \na, 5, 53, \dodoi{10.1016/S1384-1076(00)00006-3}

\bibitem[{{Budden}(1952)}]{1952RSPSA.215..215B}
{Budden}, K.~G. 1952, Proceedings of the Royal Society of London Series A, 215,
  215, \dodoi{10.1098/rspa.1952.0207}

\bibitem[{{Buenzli} \& {Schmid}(2009)}]{Buenzli09}
{Buenzli}, E., \& {Schmid}, H.~M. 2009, \aap, 504, 259,
  \dodoi{10.1051/0004-6361/200911760}

\bibitem[{{Burderi} {et~al.}(2000){Burderi}, {Di Salvo}, {Robba}, {La Barbera},
  \& {Guainazzi}}]{2000ApJ...530..429B}
{Burderi}, L., {Di Salvo}, T., {Robba}, N.~R., {La Barbera}, A., \&
  {Guainazzi}, M. 2000, \apj, 530, 429, \dodoi{10.1086/308336}

\bibitem[{{Caiazzo} \& {Heyl}(2021)}]{2021MNRAS.501..109C}
{Caiazzo}, I., \& {Heyl}, J. 2021, \mnras, 501, 109,
  \dodoi{10.1093/mnras/staa3428}

\bibitem[{{Daugherty} \& {Harding}(1986)}]{1986ApJ...309..362D}
{Daugherty}, J.~K., \& {Harding}, A.~K. 1986, \apj, 309, 362,
  \dodoi{10.1086/164608}

\bibitem[{{Di Marco} {et~al.}(2022){Di Marco}, {Costa}, {Muleri}, {Soffitta},
  {Fabiani}, {La Monaca}, {Rankin}, {Xie}, {Bachetti}, {Baldini},
  {Baumgartner}, {Bellazzini}, {Brez}, {Castellano}, {Del Monte}, {Di Lalla},
  {Ferrazzoli}, {Latronico}, {Maldera}, {Manfreda}, {O'Dell}, {Perri},
  {Pesce-Rollins}, {Puccetti}, {Ramsey}, {Ratheesh}, {Sgr{\`o}}, {Spandre},
  {Tennant}, {Tobia}, {Trois}, \& {Weisskopf}}]{Di_Marco_2022}
{Di Marco}, A., {Costa}, E., {Muleri}, F., {et~al.} 2022, \aj, 163, 170,
  \dodoi{10.3847/1538-3881/ac51c9}

\bibitem[{{Doroshenko} {et~al.}(2022){Doroshenko}, {Poutanen}, {Tsygankov},
  {Suleimanov}, {Bachetti}, {Caiazzo}, {Costa}, {Di Marco}, {Heyl}, {La
  Monaca}, {Muleri}, {Mushtukov}, {Pavlov}, {Ramsey}, {Rankin}, {Santangelo},
  {Soffitta}, {Staubert}, {Weisskopf}, {Zane}, {Agudo}, {Antonelli}, {Baldini},
  {Baumgartner}, {Bellazzini}, {Bianchi}, {Bongiorno}, {Bonino}, {Brez},
  {Bucciantini}, {Capitanio}, {Castellano}, {Cavazzuti}, {Ciprini}, {De Rosa},
  {Del Monte}, {Di Gesu}, {Di Lalla}, {Donnarumma}, {Dovciak}, {Ehlert},
  {Enoto}, {Evangelista}, {Fabiani}, {Ferrazzoli}, {Garcia}, {Gunji},
  {Hayashida}, {Iwakiri}, {Jorstad}, {Karas}, {Kitaguchi}, {Kolodziejczak},
  {Krawczynski}, {Latronico}, {Liodakis}, {Maldera}, {Manfreda}, {Marin},
  {Marinucci}, {Marscher}, {Marshall}, {Massaro}, {Matt}, {Mitsuishi},
  {Mizuno}, {Ng}, {O'Dell}, {Omodei}, {Oppedisano}, {Papitto}, {Peirson},
  {Perri}, {Pesce-Rollins}, {Petrucci}, {Pilia}, {Possenti}, {Puccetti},
  {Ratheesh}, {Romani}, {Sgro}, {Slane}, {Spandre}, {Sunyaev}, {Tamagawa},
  {Tavecchio}, {Taverna}, {Tawara}, {Tennant}, {Thomas}, {Tombesi}, {Trois},
  {Turolla}, {Vink}, {Wu}, \& {Xie}}]{Doroshenko2022}
{Doroshenko}, V., {Poutanen}, J., {Tsygankov}, S., {et~al.} 2022, arXiv
  e-prints, arXiv:2206.07138.
\newblock \doarXiv{2206.07138}

\bibitem[{{Foreman-Mackey} {et~al.}(2013){Foreman-Mackey}, {Hogg}, {Lang}, \&
  {Goodman}}]{2013PASP..125..306F}
{Foreman-Mackey}, D., {Hogg}, D.~W., {Lang}, D., \& {Goodman}, J. 2013, \pasp,
  125, 306, \dodoi{10.1086/670067}

\bibitem[{{Ghosh} \& {Lamb}(1979)}]{1979ApJ...234..296G}
{Ghosh}, P., \& {Lamb}, F.~K. 1979, \apj, 234, 296, \dodoi{10.1086/157498}

\bibitem[{{Giacconi} {et~al.}(1971){Giacconi}, {Gursky}, {Kellogg}, {Schreier},
  \& {Tananbaum}}]{1971ApJ...167L..67G}
{Giacconi}, R., {Gursky}, H., {Kellogg}, E., {Schreier}, E., \& {Tananbaum}, H.
  1971, \apjl, 167, L67, \dodoi{10.1086/180762}

\bibitem[{{Gnedin} \& {Pavlov}(1974)}]{1974JETP...38..903G}
{Gnedin}, Y.~N., \& {Pavlov}, G.~G. 1974, Soviet Journal of Experimental and
  Theoretical Physics, 38, 903

\bibitem[{{Gnedin} {et~al.}(1978){Gnedin}, {Pavlov}, \&
  {Shibanov}}]{1978PAZh....4..214G}
{Gnedin}, Y.~N., {Pavlov}, G.~G., \& {Shibanov}, I.~A. 1978, Pisma v
  Astronomicheskii Zhurnal, 4, 214

\bibitem[{{Gnedin} \& {Sunyaev}(1974)}]{Gnedin74}
{Gnedin}, Y.~N., \& {Sunyaev}, R.~A. 1974, \aap, 36, 379

\bibitem[{{Gonz{\'a}lez-Caniulef} {et~al.}(2022){Gonz{\'a}lez-Caniulef},
  {Caiazzo}, \& {Heyl}}]{unbinned}
{Gonz{\'a}lez-Caniulef}, D., {Caiazzo}, I., \& {Heyl}, J. 2022, \mnras,
  submitted, arXiv:2204.00140.
\newblock \doarXiv{2204.00140}

\bibitem[{{Gonz{\'a}lez-Caniulef} {et~al.}(2019){Gonz{\'a}lez-Caniulef},
  {Zane}, {Turolla}, \& {Wu}}]{2019MNRAS.483..599G}
{Gonz{\'a}lez-Caniulef}, D., {Zane}, S., {Turolla}, R., \& {Wu}, K. 2019,
  \mnras, 483, 599, \dodoi{10.1093/mnras/sty3159}

\bibitem[{{Hainich} {et~al.}(2019){Hainich}, {Ramachandran}, {Shenar},
  {Sander}, {Todt}, {Gruner}, {Oskinova}, \& {Hamann}}]{2019A&A...621A..85H}
{Hainich}, R., {Ramachandran}, V., {Shenar}, T., {et~al.} 2019, \aap, 621, A85,
  \dodoi{10.1051/0004-6361/201833787}

\bibitem[{{Harding} \& {Lai}(2006)}]{2006RPPh...69.2631H}
{Harding}, A.~K., \& {Lai}, D. 2006, Reports on Progress in Physics, 69, 2631,
  \dodoi{10.1088/0034-4885/69/9/R03}

\bibitem[{{Heindl} \& {Chakrabarty}(1999)}]{1999hxra.conf...25H}
{Heindl}, W.~A., \& {Chakrabarty}, D. 1999, in Highlights in X-ray Astronomy,
  MPE rept. 272, ed. B.~{Aschenbach} \& M.~J. {Freyberg} (Garching: MPI für
  Extraterrestrische Physik), 25

\bibitem[{{Heyl} \& {Caiazzo}(2018)}]{2018Galax...6...76H}
{Heyl}, J., \& {Caiazzo}, I. 2018, Galaxies, 6, 76,
  \dodoi{10.3390/galaxies6030076}

\bibitem[{{Heyl} \& {Shaviv}(2002)}]{2002PhRvD..66b3002H}
{Heyl}, J.~S., \& {Shaviv}, N.~J. 2002, \prd, 66, 023002,
  \dodoi{10.1103/PhysRevD.66.023002}

\bibitem[{{Ji} {et~al.}(2019){Ji}, {Staubert}, {Ducci}, {Santangelo}, {Zhang},
  \& {Chang}}]{2019MNRAS.484.3797J}
{Ji}, L., {Staubert}, R., {Ducci}, L., {et~al.} 2019, \mnras, 484, 3797,
  \dodoi{10.1093/mnras/stz264}

\bibitem[{{Kaminker} {et~al.}(1982){Kaminker}, {Pavlov}, \&
  {Shibanov}}]{1982Ap&SS..86..249K}
{Kaminker}, A.~D., {Pavlov}, G.~G., \& {Shibanov}, I.~A. 1982, \apss, 86, 249,
  \dodoi{10.1007/BF00683336}

\bibitem[{{Kii}(1987)}]{1987PASJ...39..781K}
{Kii}, T. 1987, \pasj, 39, 781

\bibitem[{{Kii} {et~al.}(1986){Kii}, {Hayakawa}, {Nagase}, {Ikegami}, \&
  {Kawai}}]{1986PASJ...38..751K}
{Kii}, T., {Hayakawa}, S., {Nagase}, F., {Ikegami}, T., \& {Kawai}, N. 1986,
  \pasj, 38, 751

\bibitem[{{Kislat} {et~al.}(2015){Kislat}, {Clark}, {Beilicke}, \&
  {Krawczynski}}]{2015APh....68...45K}
{Kislat}, F., {Clark}, B., {Beilicke}, M., \& {Krawczynski}, H. 2015,
  Astroparticle Physics, 68, 45, \dodoi{10.1016/j.astropartphys.2015.02.007}

\bibitem[{{Kraus} {et~al.}(1996){Kraus}, {Blum}, {Schulte}, {Ruder}, \&
  {Meszaros}}]{1996ApJ...467..794K}
{Kraus}, U., {Blum}, S., {Schulte}, J., {Ruder}, H., \& {Meszaros}, P. 1996,
  \apj, 467, 794, \dodoi{10.1086/177653}

\bibitem[{{Krzeminski}(1974)}]{1974ApJ...192L.135K}
{Krzeminski}, W. 1974, \apjl, 192, L135, \dodoi{10.1086/181609}

\bibitem[{{Kylafis} {et~al.}(2021){Kylafis}, {Tr{\"u}mper}, \&
  {Loudas}}]{2021A&A...655A..39K}
{Kylafis}, N.~D., {Tr{\"u}mper}, J.~E., \& {Loudas}, N.~A. 2021, \aap, 655,
  A39, \dodoi{10.1051/0004-6361/202039361}

\bibitem[{{Lai} \& {Ho}(2003)}]{2003PhRvL..91g1101L}
{Lai}, D., \& {Ho}, W.~C. 2003, \prl, 91, 071101,
  \dodoi{10.1103/PhysRevLett.91.071101}

\bibitem[{{Lai} \& {Ho}(2002)}]{2002ApJ...566..373L}
{Lai}, D., \& {Ho}, W. C.~G. 2002, \apj, 566, 373, \dodoi{10.1086/338074}

\bibitem[{{Lyubarskii} \& {Syunyaev}(1988)}]{1988SvAL...14..390L}
{Lyubarskii}, Y.~E., \& {Syunyaev}, R.~A. 1988, Soviet Astronomy Letters, 14,
  390

\bibitem[{{Marshall}(2021)}]{2021AJ....162..134M}
{Marshall}, H.~L. 2021, \aj, 162, 134, \dodoi{10.3847/1538-3881/ac173d}

\bibitem[{{Matt}(1993)}]{Matt93}
{Matt}, G. 1993, \mnras, 260, 663, \dodoi{10.1093/mnras/260.3.663}

\bibitem[{{Meszaros} \& {Nagel}(1985{\natexlab{a}})}]{1985ApJ...298..147M}
{Meszaros}, P., \& {Nagel}, W. 1985{\natexlab{a}}, \apj, 298, 147,
  \dodoi{10.1086/163594}

\bibitem[{{Meszaros} \& {Nagel}(1985{\natexlab{b}})}]{1985ApJ...299..138M}
---. 1985{\natexlab{b}}, \apj, 299, 138, \dodoi{10.1086/163687}

\bibitem[{{Meszaros} {et~al.}(1988){Meszaros}, {Novick}, {Szentgyorgyi},
  {Chanan}, \& {Weisskopf}}]{1988ApJ...324.1056M}
{Meszaros}, P., {Novick}, R., {Szentgyorgyi}, A., {Chanan}, G.~A., \&
  {Weisskopf}, M.~C. 1988, \apj, 324, 1056, \dodoi{10.1086/165962}

\bibitem[{{Mushtukov} \& {Tsygankov}(2022)}]{2022arXiv220414185M}
{Mushtukov}, A., \& {Tsygankov}, S. 2022, arXiv e-prints, arXiv:2204.14185.
\newblock \doarXiv{2204.14185}

\bibitem[{{Mushtukov} {et~al.}(2022){Mushtukov}, {Markozov}, {Suleimanov},
  {Nagirner}, {Kaminker}, {Potekhin}, \& {Portegies
  Zwart}}]{2022PhRvD.105j3027M}
{Mushtukov}, A.~A., {Markozov}, I.~D., {Suleimanov}, V.~F., {et~al.} 2022,
  \prd, 105, 103027, \dodoi{10.1103/PhysRevD.105.103027}

\bibitem[{{Mushtukov} {et~al.}(2017){Mushtukov}, {Suleimanov}, {Tsygankov}, \&
  {Ingram}}]{2017MNRAS.467.1202M}
{Mushtukov}, A.~A., {Suleimanov}, V.~F., {Tsygankov}, S.~S., \& {Ingram}, A.
  2017, \mnras, 467, 1202, \dodoi{10.1093/mnras/stx141}

\bibitem[{{Mushtukov} {et~al.}(2021){Mushtukov}, {Suleimanov}, {Tsygankov}, \&
  {Portegies Zwart}}]{2021MNRAS.503.5193M}
{Mushtukov}, A.~A., {Suleimanov}, V.~F., {Tsygankov}, S.~S., \& {Portegies
  Zwart}, S. 2021, \mnras, 503, 5193, \dodoi{10.1093/mnras/stab811}

\bibitem[{{Mushtukov} {et~al.}(2015){Mushtukov}, {Suleimanov}, {Tsygankov}, \&
  {Poutanen}}]{2015MNRAS.447.1847M}
{Mushtukov}, A.~A., {Suleimanov}, V.~F., {Tsygankov}, S.~S., \& {Poutanen}, J.
  2015, \mnras, 447, 1847, \dodoi{10.1093/mnras/stu2484}

\bibitem[{{Nagase} {et~al.}(1992){Nagase}, {Corbet}, {Day}, {Inoue},
  {Takeshima}, {Yoshida}, \& {Mihara}}]{1992ApJ...396..147N}
{Nagase}, F., {Corbet}, R.~H.~D., {Day}, C.~S.~R., {et~al.} 1992, \apj, 396,
  147, \dodoi{10.1086/171705}

\bibitem[{{Nagel}(1981{\natexlab{a}})}]{1981ApJ...251..278N}
{Nagel}, W. 1981{\natexlab{a}}, \apj, 251, 278, \dodoi{10.1086/159463}

\bibitem[{{Nagel}(1981{\natexlab{b}})}]{1981ApJ...251..288N}
---. 1981{\natexlab{b}}, \apj, 251, 288, \dodoi{10.1086/159464}

\bibitem[{{Paczy{\'n}ski}(1971)}]{Paczynski71}
{Paczy{\'n}ski}, B. 1971, \araa, 9, 183,
  \dodoi{10.1146/annurev.aa.09.090171.001151}

\bibitem[{{Pavlov} \& {Shibanov}(1979)}]{1979JETP...49..741P}
{Pavlov}, G.~G., \& {Shibanov}, Y.~A. 1979, Soviet Journal of Experimental and
  Theoretical Physics, 49, 741

\bibitem[{{Postnov} {et~al.}(2015){Postnov}, {Gornostaev}, {Klochkov},
  {Laplace}, {Lukin}, \& {Shakura}}]{2015MNRAS.452.1601P}
{Postnov}, K.~A., {Gornostaev}, M.~I., {Klochkov}, D., {et~al.} 2015, \mnras,
  452, 1601, \dodoi{10.1093/mnras/stv1393}

\bibitem[{{Poutanen}(2020)}]{Poutanen20RVM}
{Poutanen}, J. 2020, \aap, 641, A166, \dodoi{10.1051/0004-6361/202038689}

\bibitem[{{Poutanen} {et~al.}(2013){Poutanen}, {Mushtukov}, {Suleimanov},
  {Tsygankov}, {Nagirner}, {Doroshenko}, \& {Lutovinov}}]{Poutanen13}
{Poutanen}, J., {Mushtukov}, A.~A., {Suleimanov}, V.~F., {et~al.} 2013, \apj,
  777, 115, \dodoi{10.1088/0004-637X/777/2/115}

\bibitem[{{Poutanen} {et~al.}(1996){Poutanen}, {Nagendra}, \&
  {Svensson}}]{Poutanen96}
{Poutanen}, J., {Nagendra}, K.~N., \& {Svensson}, R. 1996, \mnras, 283, 892,
  \dodoi{10.1093/mnras/283.3.892}

\bibitem[{{Radhakrishnan} \& {Cooke}(1969)}]{1969ApL.....3..225R}
{Radhakrishnan}, V., \& {Cooke}, D.~J. 1969, \aplett, 3, 225

\bibitem[{{Raichur} \& {Paul}(2010)}]{2010MNRAS.401.1532R}
{Raichur}, H., \& {Paul}, B. 2010, \mnras, 401, 1532,
  \dodoi{10.1111/j.1365-2966.2009.15778.x}

\bibitem[{{Sanjurjo-Ferr{\'\i}n} {et~al.}(2021){Sanjurjo-Ferr{\'\i}n},
  {Torrej{\'o}n}, {Postnov}, {Oskinova}, {Rodes-Roca}, \&
  {Bernabeu}}]{Sanjurjo-Ferrin21}
{Sanjurjo-Ferr{\'\i}n}, G., {Torrej{\'o}n}, J.~M., {Postnov}, K., {et~al.}
  2021, \mnras, 501, 5892, \dodoi{10.1093/mnras/staa3953}

\bibitem[{{Santangelo} {et~al.}(1998){Santangelo}, {del Sordo}, {Segreto}, {dal
  Fiume}, {Orlandini}, \& {Piraino}}]{1998A&A...340L..55S}
{Santangelo}, A., {del Sordo}, S., {Segreto}, A., {et~al.} 1998, \aap, 340, L55

\bibitem[{{Schreier} {et~al.}(1972){Schreier}, {Levinson}, {Gursky}, {Kellogg},
  {Tananbaum}, \& {Giacconi}}]{1972ApJ...172L..79S}
{Schreier}, E., {Levinson}, R., {Gursky}, H., {et~al.} 1972, \apjl, 172, L79,
  \dodoi{10.1086/180896}

\bibitem[{{Silver} {et~al.}(1979){Silver}, {Weisskopf}, {Kestenbaum}, {Long},
  {Novick}, \& {Wolff}}]{1979ApJ...232..248S}
{Silver}, E.~H., {Weisskopf}, M.~C., {Kestenbaum}, H.~L., {et~al.} 1979, \apj,
  232, 248, \dodoi{10.1086/157283}

\bibitem[{{Siuniaev}(1976)}]{1976SvAL....2..111S}
{Siuniaev}, R.~A. 1976, Soviet Astronomy Letters, 2, 111

\bibitem[{{Sobolev}(1963)}]{Sobolev63}
{Sobolev}, V.~V. 1963, {A Treatise on Radiative Transfer} (Princeton: Van
  Nostrand)

\bibitem[{{Soffitta} {et~al.}(2021){Soffitta}, {Baldini}, {Bellazzini},
  {Costa}, {Latronico}, {Muleri}, {Del Monte}, {Fabiani}, {Minuti}, {Pinchera},
  {Sgro'}, {Spandre}, {Trois}, {Amici}, {Andersson}, {Attina'}, {Bachetti},
  {Barbanera}, {Borotto}, {Brez}, {Brienza}, {Caporale}, {Cardelli},
  {Carpentiero}, {Castellano}, {Castronuovo}, {Cavalli}, {Cavazzuti},
  {Ceccanti}, {Centrone}, {Ciprini}, {Citraro}, {D'Amico}, {D'Alba}, {Di
  Cosimo}, {Di Lalla}, {Di Marco}, {Di Persio}, {Donnarumma}, {Evangelista},
  {Ferrazzoli}, {Hayato}, {Kitaguchi}, {La Monaca}, {Lefevre}, {Loffredo},
  {Lorenzi}, {Lucchesi}, {Magazzu}, {Maldera}, {Manfreda}, {Mangraviti},
  {Marengo}, {Matt}, {Mereu}, {Morbidini}, {Mosti}, {Nakano}, {Nasimi},
  {Negri}, {Nenonen}, {Nuti}, {Orsini}, {Perri}, {Pesce-Rollins}, {Piazzolla},
  {Pilia}, {Profeti}, {Puccetti}, {Rankin}, {Ratheesh}, {Rubini}, {Santoli},
  {Sarra}, {Scalise}, {Sciortino}, {Tamagawa}, {Tardiola}, {Tobia},
  {Vimercati}, \& {Xie}}]{2021AJ....162..208S}
{Soffitta}, P., {Baldini}, L., {Bellazzini}, R., {et~al.} 2021, \aj, 162, 208,
  \dodoi{10.3847/1538-3881/ac19b0}

\bibitem[{{Sokolova-Lapa} {et~al.}(2021){Sokolova-Lapa}, {Gornostaev}, {Wilms},
  {Ballhausen}, {Falkner}, {Postnov}, {Thalhammer}, {F{\"u}rst}, {Garc{\'\i}a},
  {Shakura}, {Becker}, {Wolff}, {Pottschmidt}, {H{\"a}rer}, \&
  {Malacaria}}]{2021A&A...651A..12S}
{Sokolova-Lapa}, E., {Gornostaev}, M., {Wilms}, J., {et~al.} 2021, \aap, 651,
  A12, \dodoi{10.1051/0004-6361/202040228}

\bibitem[{{Strohmayer}(2017)}]{2017ApJ...838...72S}
{Strohmayer}, T.~E. 2017, \apj, 838, 72, \dodoi{10.3847/1538-4357/aa643d}

\bibitem[{{Suchy} {et~al.}(2008){Suchy}, {Pottschmidt}, {Wilms}, {Kreykenbohm},
  {Sch{\"o}nherr}, {Kretschmar}, {McBride}, {Caballero}, {Rothschild}, \&
  {Grinberg}}]{2008ApJ...675.1487S}
{Suchy}, S., {Pottschmidt}, K., {Wilms}, J., {et~al.} 2008, \apj, 675, 1487,
  \dodoi{10.1086/527042}

\bibitem[{{Suleimanov} {et~al.}(2010){Suleimanov}, {Pavlov}, \&
  {Werner}}]{2010ApJ...714..630S}
{Suleimanov}, V.~F., {Pavlov}, G.~G., \& {Werner}, K. 2010, \apj, 714, 630,
  \dodoi{10.1088/0004-637X/714/1/630}

\bibitem[{{Tjemkes} {et~al.}(1986){Tjemkes}, {Zuiderwijk}, \& {van
  Paradijs}}]{1986A&A...154...77T}
{Tjemkes}, S.~A., {Zuiderwijk}, E.~J., \& {van Paradijs}, J. 1986, \aap, 154,
  77

\bibitem[{{Tomar} {et~al.}(2021){Tomar}, {Pradhan}, \&
  {Paul}}]{2021MNRAS.500.3454T}
{Tomar}, G., {Pradhan}, P., \& {Paul}, B. 2021, \mnras, 500, 3454,
  \dodoi{10.1093/mnras/staa3477}

\bibitem[{{Verbunt}(1999)}]{Verbunt99}
{Verbunt}, F. 1999, in Astronomical Society of the Pacific Conference Series,
  Vol. 160, Astrophysical Discs - an EC Summer School, ed. J.~A. {Sellwood} \&
  J.~{Goodman}, 21

\bibitem[{{Weisskopf} {et~al.}(2022){Weisskopf}, {Soffitta}, {Baldini},
  {Ramsey}, {O'Dell}, {Romani}, {Matt}, {Deininger}, {Baumgartner},
  {Bellazzini}, {Costa}, {Kolodziejczak}, {Latronico}, {Marshall}, {Muleri},
  {Bongiorno}, {Tennant}, {Bucciantini}, {Dovciak}, {Marin}, {Marscher},
  {Poutanen}, {Slane}, {Turolla}, {Kalinowski}, {Di Marco}, {Fabiani},
  {Minuti}, {La Monaca}, {Pinchera}, {Rankin}, {Sgro'}, {Trois}, {Xie},
  {Alexander}, {Allen}, {Amici}, {Andersen}, {Antonelli}, {Antoniak},
  {Attina'}, {Barbanera}, {Bachetti}, {Baggett}, {Bladt}, {Brez}, {Bonino},
  {Boree}, {Borotto}, {Breeding}, {Brienza}, {Bygott}, {Caporale}, {Cardelli},
  {Carpentiero}, {Castellano}, {Castronuovo}, {Cavalli}, {Cavazzuti},
  {Ceccanti}, {Centrone}, {Citraro}, {D' Amico}, {D'Alba}, {Di Gesu}, {Del
  Monte}, {Dietz}, {Di Lalla}, {Di Persio}, {Dolan}, {Donnarumma},
  {Evangelista}, {Ferrant}, {Ferrazzoli}, {Ferrie}, {Footdale}, {Forsyth},
  {Foster}, {Garelick}, {Gunji}, {Gurnee}, {Head}, {Hibbard}, {Johnson},
  {Kelly}, {Kilaru}, {Lefevre}, {Le Roy}, {Loffredo}, {Lorenzi}, {Lucchesi},
  {Maddox}, {Magazzu}, {Maldera}, {Manfreda}, {Mangraviti}, {Marengo},
  {Marrocchesi}, {Massaro}, {Mauger}, {McCracken}, {McEachen}, {Mize}, {Mereu},
  {Mitchell}, {Mitsuishi}, {Morbidini}, {Mosti}, {Nasimi}, {Negri}, {Negro},
  {Nguyen}, {Nitschke}, {Nuti}, {Onizuka}, {Oppedisano}, {Orsini}, {Osborne},
  {Pacheco}, {Paggi}, {Painter}, {Pavelitz}, {Pentz}, {Piazzolla}, {Perri},
  {Pesce-Rollins}, {Peterson}, {Pilia}, {Profeti}, {Puccetti}, {Ranganathan},
  {Ratheesh}, {Reedy}, {Root}, {Rubini}, {Ruswick}, {Sanchez}, {Sarra},
  {Santoli}, {Scalise}, {Sciortino}, {Schroeder}, {Seek}, {Sosdian}, {Spandre},
  {Speegle}, {Tamagawa}, {Tardiola}, {Tobia}, {Thomas}, {Valerie}, {Vimercati},
  {Walden}, {Weddendorf}, {Wedmore}, {Welch}, {Zanetti}, \&
  {Zanetti}}]{Weisskopf2022}
{Weisskopf}, M.~C., {Soffitta}, P., {Baldini}, L., {et~al.} 2022, J. Astron.
  Telesc. Instrum. Syst., 8, 026002, \dodoi{10.1117/1.JATIS.8.2.026002}

\bibitem[{{Wilms} {et~al.}(2000){Wilms}, {Allen}, \& {McCray}}]{Wilms2000}
{Wilms}, J., {Allen}, A., \& {McCray}, R. 2000, \apj, 542, 914,
  \dodoi{10.1086/317016}

\end{thebibliography}
\bibliographystyle{aasjournal}



\end{document}